\documentclass[10pt,twocolumn]{article}
\usepackage{graphicx}
\usepackage{fancyhdr}
\usepackage{makeidx}
\usepackage{amssymb,amsmath}
\usepackage{enumerate}
\newcommand{\bd}{\begin{document}}
\newcommand{\ed}{\end{document}}
\newcommand{\bc}{\begin{center}}
\newcommand{\ec}{\end{center}}
\newcommand{\vs}{\vspace}
\newcommand{\hs}{\hspace}
\newcommand{\beq}{\begin{equation}}
\newcommand{\eeq}{\end{equation}}
\newcommand{\beqs}{\begin{eqn*}}
\newcommand{\eeqs}{\end{eqn*}}
\newcommand{\bq}{\begin{quote}}
\newcommand{\eq}{\end{quote}}
\newcommand{\lb}{\linebreak}
\newcommand{\mb}{\makebox}
\newcommand{\fb}{\framebox}
\newcommand{\mc}{\multicolumn}
\newcommand{\ben}{\begin{enumerate}}
\newcommand{\een}{\end{enumerate}}
\newcommand{\bit}{\begin{itemize}}
\newcommand{\eit}{\end{itemize}}
\newcommand{\ov}{\overline}
\newcommand{\un}{\underline}
\newcommand{\lt}{\left}
\newcommand{\rt}{\right}
\newcommand{\ba}{\begin{array}}
\newcommand{\ea}{\end{array}}
\newcommand{\beqa}{\begin{eqnarray}}
\newcommand{\eeqa}{\end{eqnarray}}
\newcommand{\beqas}{\begin{eqnarray*}}
\newcommand{\eeqas}{\end{eqnarray*}}
\newcommand{\bfg}{\begin{figure}}
\newcommand{\efg}{\end{figure}}
\newcommand{\pad}{\partial}
\newcommand{\nn}{\nonumber}
\newcommand{\la}{\leftarrow}
\newcommand{\ra}{\rightarrow}
\newcommand{\lgla}{\longleftarrow}
\newcommand{\lgra}{\longrightarrow}
\newcommand{\La}{\Leftarrow}
\newcommand{\Ra}{\Rightarrow}
\newcommand{\Lra}{\Leftrightarrow}
\newcommand{\Lgla}{\Longleftarrow}
\newcommand{\Lgra}{\Longrightarrow}
\renewcommand{\a}{\alpha}
\renewcommand{\b}{\beta}
\newcommand{\g}{\gamma}
\newcommand{\G}{\Gamma}
\renewcommand{\d}{\delta}
\newcommand{\D}{\Delta}
\newcommand{\e}{\epsilon}
\newcommand{\eps}{\epsilon}
\newcommand{\s}{\sigma}
\renewcommand{\l}{\lamda}
\newcommand{\m}{\mu}
\newcommand{\n}{\nu}
\renewcommand{\S}{\Sigma}
\newcommand{\p}{\pi}
\newcommand{\om}{\omega}
\newcommand{\Om}{\Omega}
\newcommand{\tri}{\triangle}
\newcommand{\ti}{\times}
\newcommand{\f}{\frac}
\newcommand{\ds}{\displaystyle}
\newcommand{\bm}[1]{\mb{{\boldmath $#1$}}}
\newcommand{\alter}[2]{\lt\{ \ba{ll}#1 \\ #2 \ea \rt.}
\newcommand{\alt}[4]{\lt\{ \ba{ll}#1 & \mb{if \, \,}#2 \\ #3 & \mb{}#4 \ea
    \rt.}
\newcommand{\altn}[4]{\lt\{ \ba{rl}#1 & \mb{if \, \,}#2 \\ #3 & \mb{}#4 \ea
    \rt.}
\newcommand{\altif}[4]{\lt\{ \ba{ll}#1 & \mb{if \, \,}#2 \\ #3 &
\mb{if \, \,}#4 \ea \rt.}
\newcommand{\altnif}[4]{\lt\{ \ba{rl}#1 & \mb{if \, \,}#2 \\ #3 &
\mb{if \, \,}#4 \ea \rt.}
\newcounter{algc}
\newcounter{romc}
\newcounter{Alphc}
\newcommand{\bl}{\begin{list}{{\it Step} ~\arabic{algc}~:} {\usecounter{algc}
                \setlength{\topsep}{0pt} \setlength{\itemsep}{0pt}}}
\newcommand{\el}{\end{list}}
\newcommand{\blr}{\begin{list}{~\roman{romc}~:} {\usecounter{romc}
                \setlength{\topsep}{0pt} \setlength{\itemsep}{0pt}}}
\newcommand{\elr}{\end{list}}
\newcommand{\bla}{\begin{list}{~\Alph{Alphc}~:} {\usecounter{Alphc}
                \setlength{\topsep}{0pt} \setlength{\itemsep}{0pt}}}
\newcommand{\ela}{\end{list}}

\newtheorem{theorem}{Theorem}
\begin{document}

\title{Bandstructure Effects in Ultra-Thin-Body DGFET: A Fullband Analysis}
\author{Kausik Majumdar and Navakanta Bhat\\
Department of Electrical Communication Engineering\\
Indian Institute of Science, Bangalore-560012, India\\
Email: $\{$kausik,navakant$\}$@ece.iisc.ernet.in}
\date{}
\maketitle

{\abstract
This paper discusses a few unique effects of ultra-thin-body double-gate
NMOSFET that are arising from the bandstructure of the thin film Si channel. The
bandstructure has been calculated using $10$-orbital $sp^3d^5s^*$ tight-binding
method. A number of intrinsic properties including band gap, density of states,
intrinsic carrier concentration and parabolic effective mass have been derived
from the calculated bandstructure. The spatial distributions of intrinsic carrier
concentration and $<100>$ effective mass, arising from the wavefunction of different
contributing subbands are analyzed. A self-consistent solution of
Poisson-Schrodinger coupled equation is obtained taking the full bandstructure into account,
which is then applied to an
insightful analysis of volume inversion. The spatial distribution of
carriers over the channel of a DGFET has been calculated and its effects on effective mass
and channel capacitance are discussed.
}


\section{Introduction}
The interest in Ultra-Thin-Body (UTB)
Double-Gate FET (DGFET) has grown in the recent past because of
its superior properties compared to bulk MOSFET and is being considered as one of the future
alternatives of present day bulk devices \cite{{fb87},{jf02}}.
Apart from superior gate control from both top and bottom,
the intrinsic quantum confinement provided by its unique geometric structure affects the characteristics
of UTB DGFET \cite{te03}. The different aspects of classical modeling of
DGFETs have been discussed in \cite{{yt00},{yt01},{yt04}}.
There are a number of reports on the quantum mechanical effects on DGFET,
both analytical as well as numerical \cite{{te03},{to94},{fg01},{lg02}}. To solve numerically,
one has to solve coupled Poisson-Schrodinger equations self-consistently \cite{fs72}. This is
done either by assuming some analytical $E-\bar{k}$ relationship, or by taking full bandstructure
into account. There has been
considerable amount of work on calculation of bandstructure of materials \cite{{jcs54},{djc75},{pv83},{jmj98},{tbb02},{tbb04},{sl04}}
which can be plugged into the self-consistent Poisson-Schrodinger equation numerically \cite{ar05}.

DGFET has a unique property of volume inversion which improves the transport characteristics
enormously \cite{fb87}, \cite{te03}. This can be explained with the help of quantum effects.
Another important aspect is a substantial change in transport properties depending on the
crystallographic orientation \cite{{sel03},{ap05},{js07}} which again can be analyzed
from detailed bandstructure calculation. Also, the total number of intrinsic carriers
reduces with the thinning of the channel material. This has an effect on the total
gate capacitance of DGFET \cite{{dv97},{lg06},{om07}}.

The aim of this paper is to focus on detailed analysis of some effects in UTB-DGFET which
arise entirely because of bandstructure of the channel material and  are not very apparent.
Only Silicon has been considered as the thin channel material in this work, but this can be easily
extended to other channel materials as well. The full-band structure
calculation that has been used here is based on $sp^3d^5s^*$ tight-binding
method \cite{{jmj98},{tbb02},{tbb04}}. The calculated bandstructure has then been used to predict some
intrinsic transport properties of thin film Si including band gap, density of states,
intrinsic carrier concentration and effective mass.
From this analysis, a number of features are explained
which are unique to ultra-thin film semiconductors. Following this,
Poisson-Schrodinger coupled equation is solved self-consistently taking care of the
full bandstructure. With the help of this, volume inversion phenomenon has been
critically analyzed with physical insights. This in turn throws some light on the
spatial distribution of carriers inside the DGFET channel. Taking this into account,
total channel capacitance and evolution of effective mass from source end to drain end
along the channel have been analyzed.

The rest of the paper is organized as follows: Sec. \ref{sec:bandcalc} discusses
on the details of $sp^3d^5s^*$ tight-binding method of bandstructure calculation.
The different intrinsic transport properties of ultra-thin film Silicon
have been discussed in sec. \ref{sec:intrinsic}. Poisson-Schrodinger coupled
equation has been solved self-consistently and
different related analyses have been performed in sec. \ref{sec:poiss_sch}. Finally the
paper is concluded in sec. \ref{sec:conclude}.

\section{Bandstructure Calculation}\label{sec:bandcalc}
Tight-binding method of bandstructure calculation has been
extensively studied by many researchers \cite{{jcs54},{djc75},{pv83},{jmj98},{tbb02},{tbb04},{sl04}}. In this
work, a 10-orbital $sp^3d^5s^*$ tight-binding method \cite{{jmj98},{tbb02},{tbb04}}
has been used to find bandstructure of the thin
film Si, being used as channel material. Only the onsite energies and two-center overlap
integrals of nearest neighbors have been taken into account. Spin orbit interaction has been neglected,
and thus each $k$ point in the Brillouin zone is assumed to be degenerate with two
spin states. Infinite crystal periodicity has been assumed along channel length
and width directions and thus Bloch's theorem is assumed to hold good in those directions.
However, along the thickness of the channel, the crystal is truncated to a few
monolayers, thus crystal periodicity can not be assumed in this direction.
Suppose, the thickness contains $N$ atomic monolayers. Then, the truncated crystal can
be formed by taking a basis of $N$ atoms along the thickness direction and spanning
them over the whole 2-D space. Fig. \ref{fig:monolayers} shows a $7$ monolayer thick
channel with the basis atoms shown as black dots. The channel region can be formed
by spanning the basis atoms along $x$ and $y$.
The tight-binding fitting parameters for Si, used in this work, have been
taken from \cite{{tbb02},{tbb04}}. An $N$ monolayers thick film will produce
a $10N\times10N$ tight-binding Hamiltonian \cite{ar05}. To get rid of the huge number of
surface states (whose energy eigen values often fall inside the semiconductor
band gap) caused from dangling bonds, it has been assumed that the surfaces
are completely passivated by Hydrogen. This has been achieved by artificially
increasing the onsite energies of $s$ and $p$ orbitals of the surface atoms,
as described in \cite{sl04}.
\bfg[htbp!]
\hs{-0.1in}
\includegraphics[scale=0.32, angle=-90]{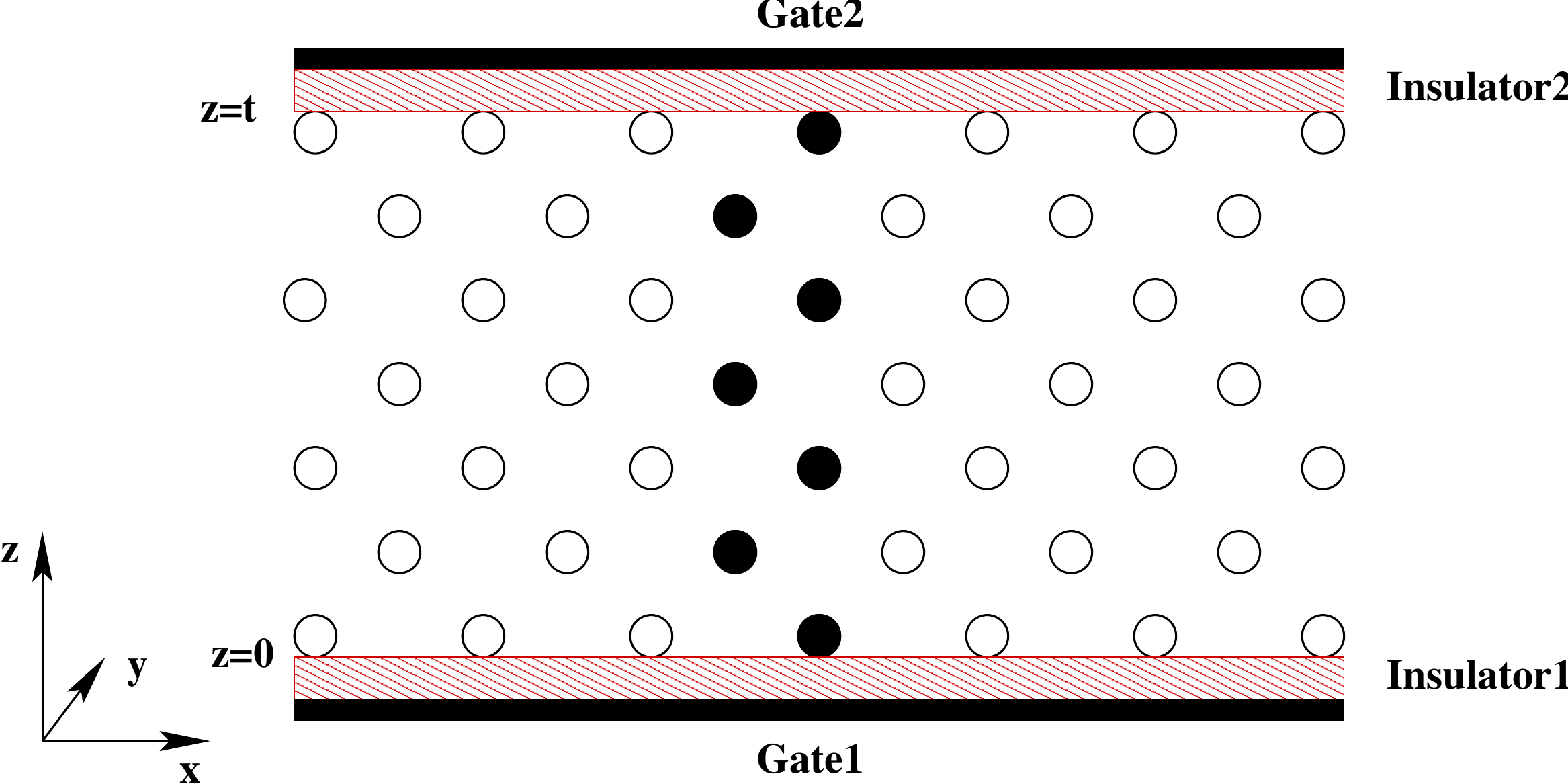}
\caption{A $7$-monolayer thick film with basis atoms (black dots).
The basis atoms can be spanned in whole $2-D$ along $x$ and $y$ to construct the
thin film.}\label{fig:monolayers}
\efg

The assumption of this method is that the electronic wave function is strictly
guided in the $x-y$ plane.  Thus, the Brillouin zone will comprise of a $2$-D $k$ space, as
opposed to a $3$-D one in bulk case. $k_z$ has been assumed to be zero
throughout this paper. The whole $2$-D Brillouin zone has been discretized
using step size of $0.05\times\frac{2\pi}{a}$ for both $k_x$ and $k_y$ where
$a$ is lattice constant (=$5.43\AA$ for Si).
In this paper, the film thickness has been referenced to the number
of monolayers (AL) in the film. An $N$ AL thick Si film translates
to a thickness of $a(N-1)/4$.
Fig. \ref{fig:bandstructure_17AL} shows energy dispersion plot
of a $17$-monolayer thick ($\sim 2.17$nm) $Si$ film over the
whole $2-D$ Brillouin zone. Only the top most valence subband
and bottom most conduction subband have been included for clarity.
Throughout this paper, the valleys occurring at $\Gamma$ point and at $\sim 0.8\frac{2\pi}{a}$
along $X$ direction are termed as $\Gamma$ valley and $X$ valley
respectively.
\bfg[htbp!]
\vs{-1.7in}
\hs{-0.8in}
\includegraphics[scale=0.6]{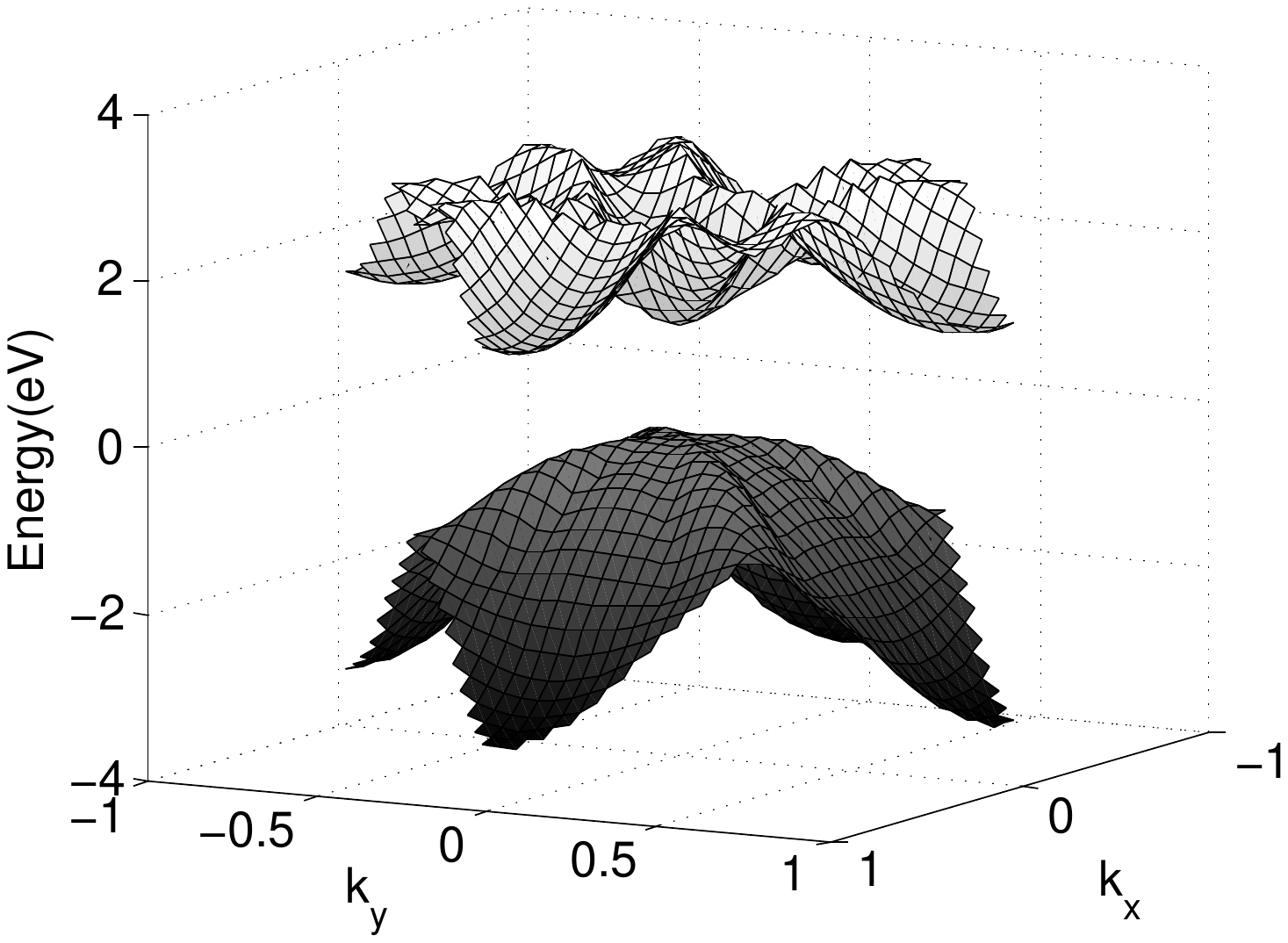}
\vs{-1.9in}
\caption{$E-k$ relationship of top most valence subband and bottom most
conduction subband over the whole $2-D$ Brillouin zone
of a $17$ monolayer thick Si film. The conduction band minimum
occurs at $\Gamma$ point. The $X$ valley is $4$ fold degenerate.}\label{fig:bandstructure_17AL}
\efg
\section{Intrinsic Properties of Thin Film Si}\label{sec:intrinsic}
In this section, the variations of different intrinsic electrical properties of thin film
as a function of film thickness have been derived from the bandstructure
calculation, described in the previous section.
\subsection{Bandgap and Density of States}
It has been well established in literature, by both theory and experiments, that
in the nano-scale, bandgap of semiconductors is a function of size of the material.
As size reduces, the bandgap of
the material increases. Fig. \ref{fig:Si_gap} shows how the $\Gamma$ gap
and $X$ gap of a Si film are changing as a function of the film thickness.
\bfg[htbp!]
\vs{-1.5in}
\hs{-0.5in}
\includegraphics[scale=0.5]{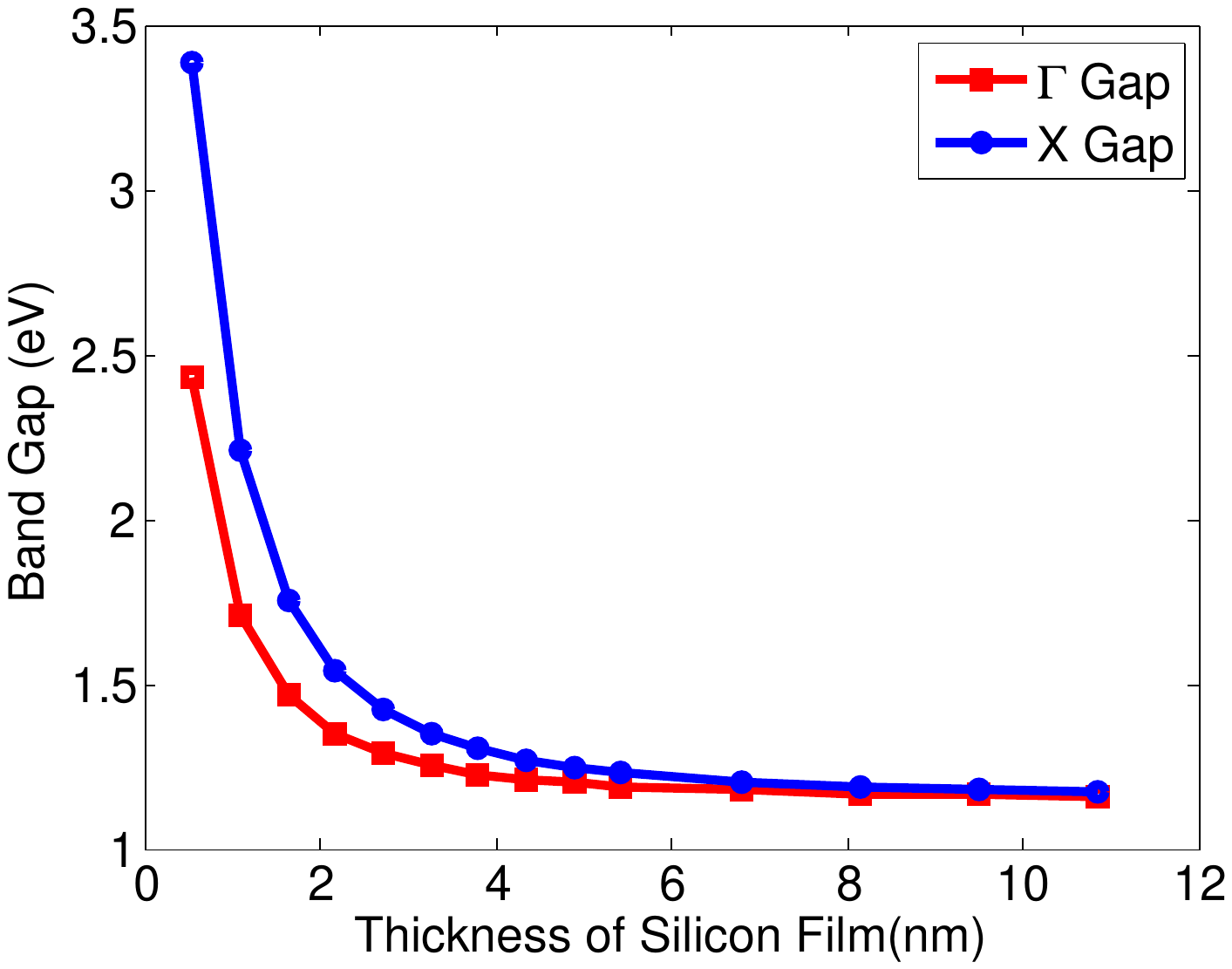}
\vs{-1.7in}
\caption{Calculated $\Gamma$ and $X$ gap of Silicon thin film as a function of
film thickness. For sufficiently small thickness, direct $\Gamma$
gap is quite larger compared to the next gap occurring at $X$ valley.}\label{fig:Si_gap}
\efg
One should note that, for a sufficiently thin film,
as opposed to the bulk case, the conduction band minimum occurs at
direct $\Gamma$ point, and not in the $X$ direction. Thus the electrons
will first populate the $\Gamma$ valley and hence, one can expect to see
drastic change in transport properties for a thin film Si channel compared to bulk.
As the film thickness increases, the energy difference $\Delta E_{\Gamma X}$
between $\Gamma$ and $X$ valleys decreases, and electron will start populating
both the valleys. Finally, at sufficiently large film thickness, at the bulk
limit, $X$ valley is of less energy compared to $\Gamma$ valley, and thus,
electrons will start populating only $X$ valley.
\bfg[htbp!]
\vs{-0.7in}
\hs{-0.4in}
\includegraphics[scale=0.45]{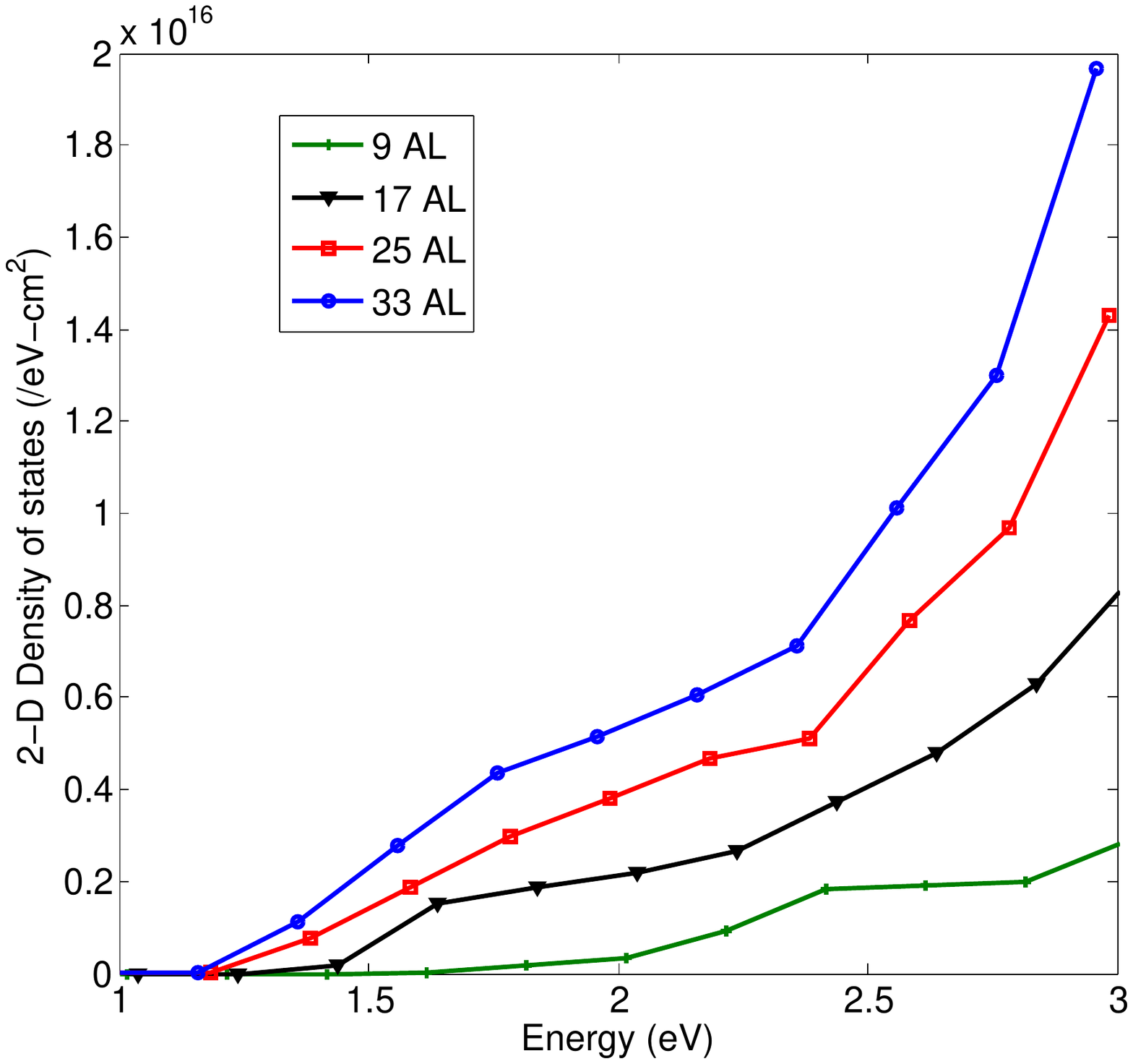}
\vs{-1.0in}
\caption{$2$-D DOS as a function of electronic energy for different
thickness values of the Si film. The energy space is discretized
by steps of $0.2$ eV.}\label{fig:Si_dos}
\efg
In Fig. \ref{fig:Si_dos}, the $2$-D density of states has been plotted
as a function of electron energy in the conduction band, for $4$ different
thickness values. In the calculation, the energy has been discretized
in steps of $0.2$eV. Just above the cut-off energy (conduction band minimum),
only $\Gamma$ valley contributes. However, as energy increases, other regions
of Brillouin zone also start contributing. As expected, density of states
for thicker film is larger.
\subsection{Intrinsic Carrier Concentration}
As it has been discussed already that bandgap increases as the size goes from bulk
to nano-scale, one expects lower intrinsic carrier concentration as the film thickness
reduces. The per unit area intrinsic electron concentration at temperature $T$ is given by
\beq
n_A = \sum_j\sum_{\bar{k}}2f(E_j^{\bar{k}})
\eeq
where the first sum is over different subband indices $j$ of conduction band
and the second sum is over all $\bar{k}$
points in the first Brillouin zone. $E_j^{\bar{k}}$ represents the energy eigenvalue at $k^{th}$ point
of $j^{th}$ subband index. The Fermi-Dirac probability $f(E_j^{\bar{k}})$ is given by
\beq
f(E_j^{\bar{k}}) = \frac{1}{1 + e^{(E_j^{\bar{k}}-\mu)/k_BT}}
\eeq
$k_B$ is the Boltzmann constant and $\mu$ is the chemical potential.
\bfg[htbp!]
\vs{-1.5in}
\hs{-0.7in}
\includegraphics[scale=0.55]{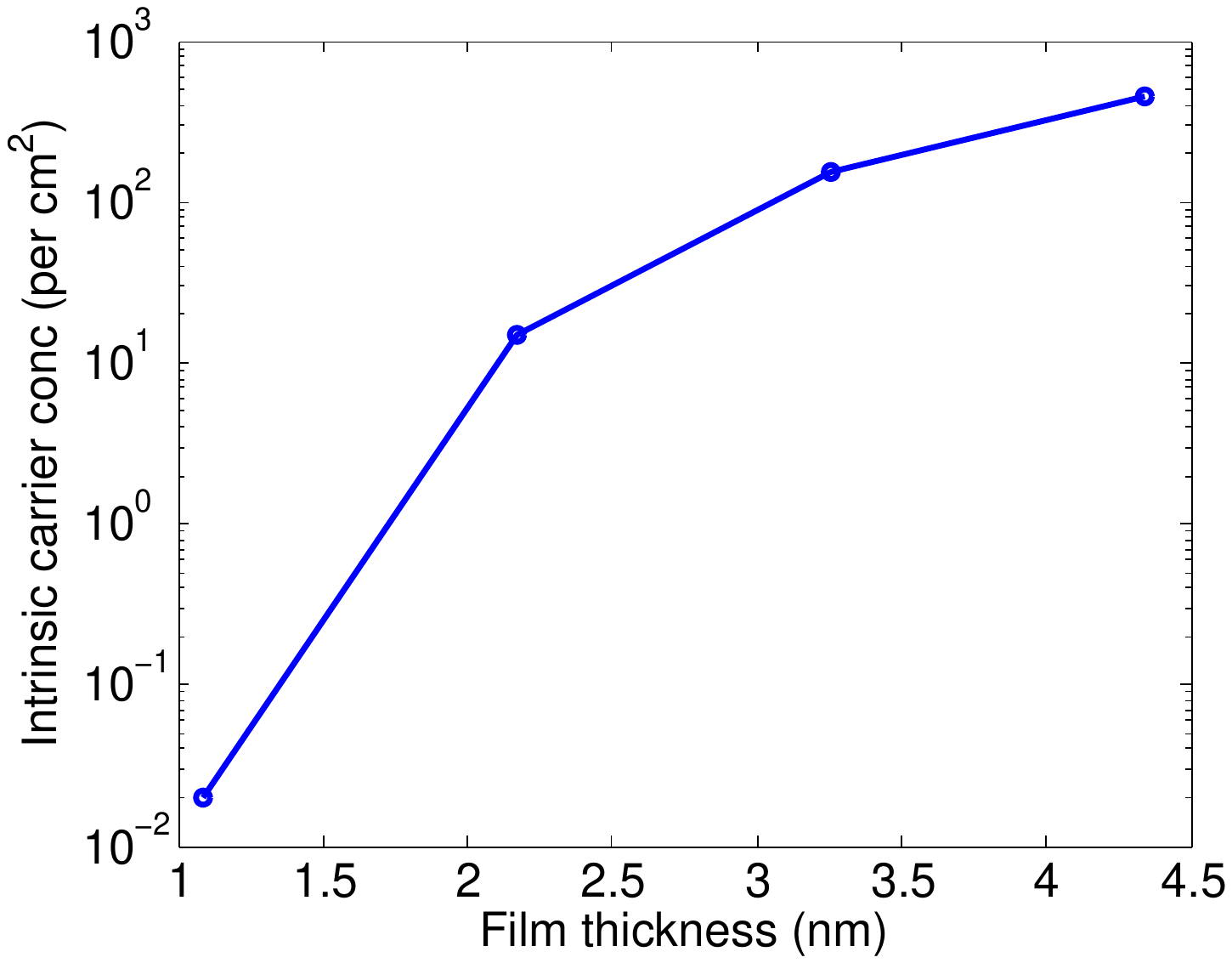}
\vs{-1.7in}
\caption{Intrinsic carrier concentration per unit area (i.e. total
number of carriers contained in the film with unity area) as a function of
Si film thickness.}\label{fig:Si_carrier_conc}
\efg
Fig. \ref{fig:Si_carrier_conc} shows that the intrinsic carrier
concentration per unit area increases with film thickness. However, apart from the reduction
in carrier concentration, another important observation is that the
carriers have a distribution along the film thickness, which peaks at the center of the film.
This is due to the spatial distribution of the wave function
of the electronic states contributing to the carrier concentration.
If the film of thickness $t$ has $N$ monolayers, then the film can be assumed
to be discretized by $N$ points, at each of which, per unit volume
carrier density is given by
\beq\label{eq:n0}
n^0(z) = \frac{N}{t}\sum_j\sum_{\bar{k}}2f(E_j^{\bar{k}})|\psi_j^{\bar{k}}(z)|^2
\eeq
where $\psi_j^{\bar{k}}(z)$ is the wave function of the electronic state $(j,\bar{k})$
at $z$.
\bfg[htbp!]
\vs{-1.5in}
\hs{-0.7in}
\includegraphics[scale=0.55]{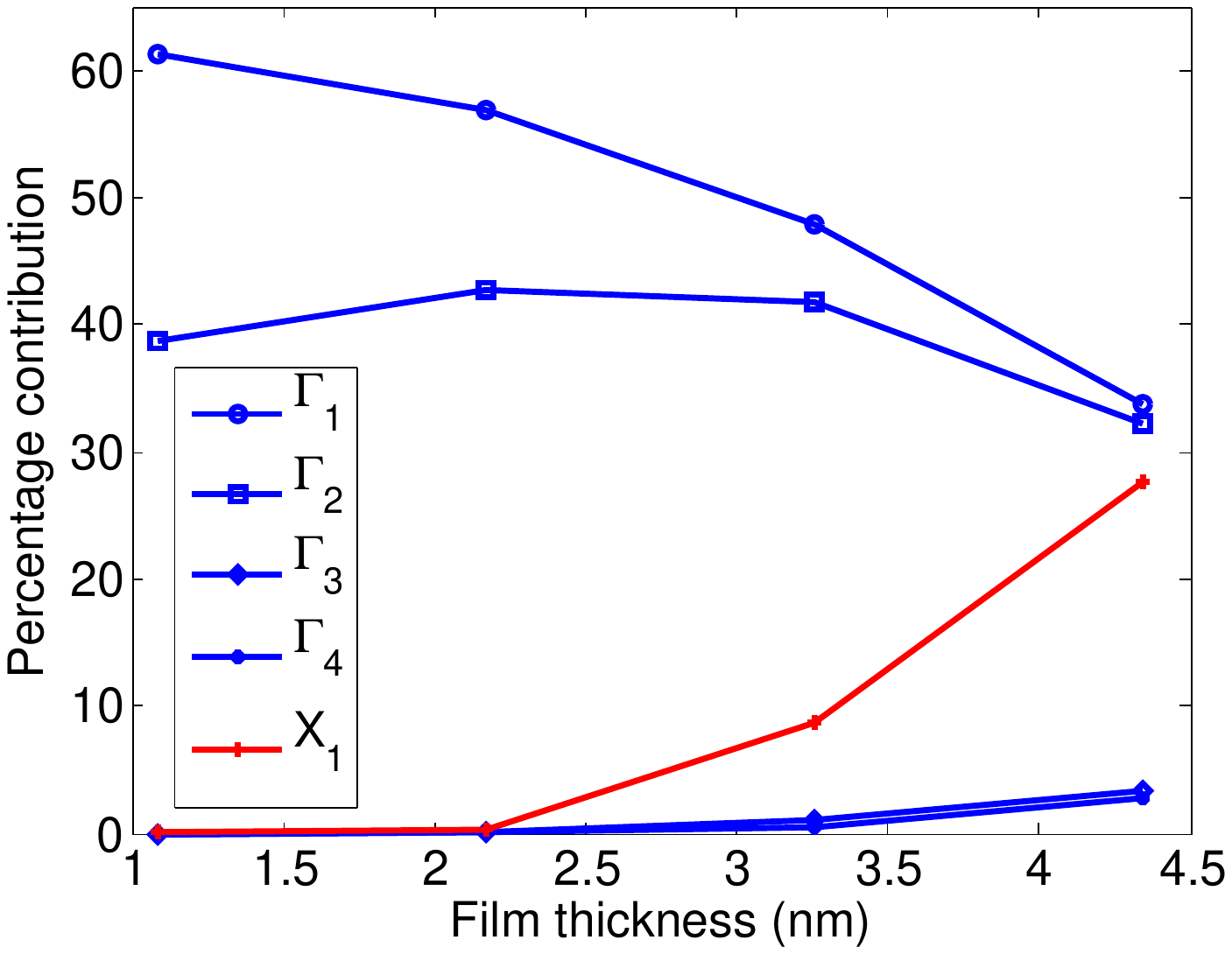}
\vs{-1.8in}
\caption{Percentage contribution to per unit area intrinsic carrier concentration
from different Si subbands lying in $\Gamma$ and $X$
valleys. For very small thickness only $\Gamma_1$ and $\Gamma_2$ contribute,
but at larger thickness, electrons start populating other valleys as well.}\label{fig:Si_subband_fraction}
\efg
Fig. \ref{fig:Si_subband_fraction} plots the fractional
contribution of different subbands to total electron concentration for
a thin film of Si. $\Gamma_i$ and $X_j$ represent the $i^{th}$ subband
of $\Gamma$ valley and $j^{th}$ subband of $X$ valley, respectively.
It is clear that, for very small thickness, only $\Gamma_1$ and $\Gamma_2$
subbands contribute, but as thickness increases, other subbands also start
contributing.
\bfg[htbp!]
\vs{-1.5in}
\hs{-0.7in}
\includegraphics[scale=0.55]{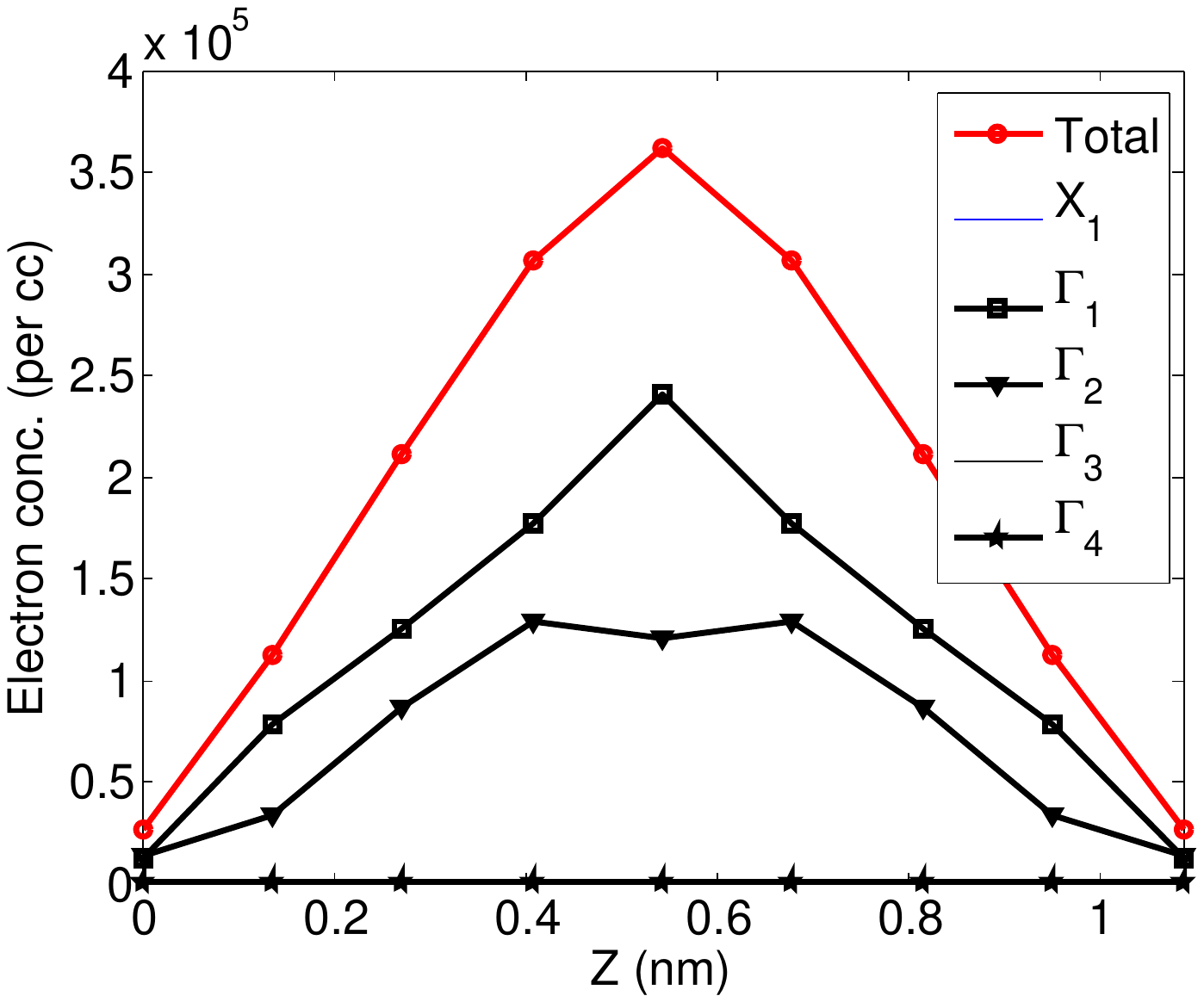}
\vs{-1.8in}
\caption{Carrier distribution along channel thickness
of different subbands lying
in $\Gamma$ and $X$ valleys for a $9$ atomic layer thick Si film.}\label{fig:conc_9AL}
\efg
\bfg[htbp!]
\vs{-1.5in}
\hs{-0.7in}
\includegraphics[scale=0.55]{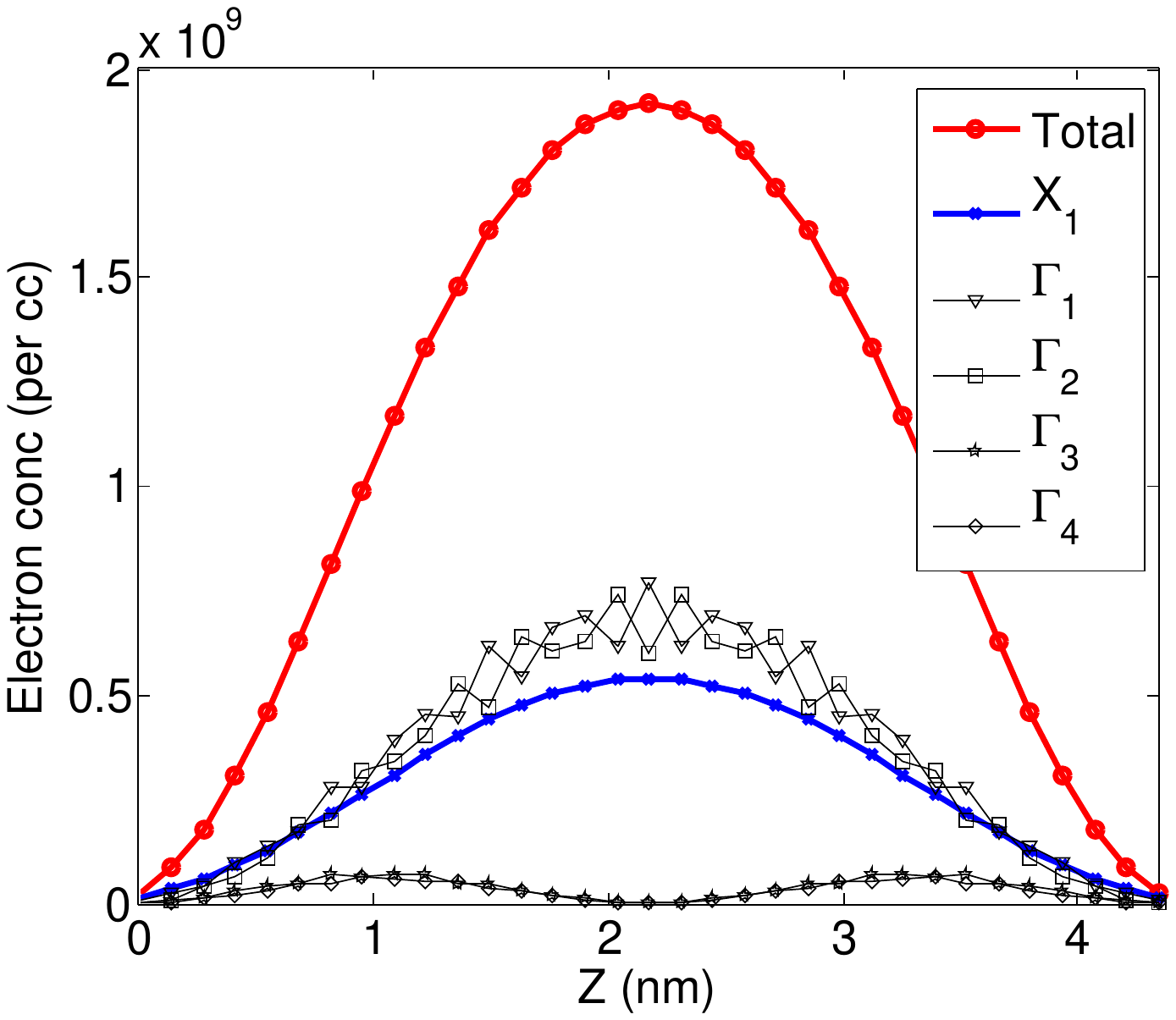}
\vs{-1.8in}
\caption{Carrier distribution along channel thickness
of different subbands lying
in $\Gamma$ and $X$ valleys for a $33$ atomic layer thick Si film.}\label{fig:conc_33AL}
\efg
In Fig. \ref{fig:conc_9AL} and \ref{fig:conc_33AL}, the spatial distribution of
intrinsic carrier concentration, coming from different subbands, have been shown,
for $9$ and $33$ monolayer thick Si film, respectively. Since for a $9$ monolayer
thick film ($\sim 1.086$nm), only $\Gamma_1$ and $\Gamma_2$ contribute, the spatial distribution
of the total electron concentration is dictated by wavefunction distribution of only these two subands.
The peak concentration comes at the middle of the film, and reduces as it approaches the
surface. However, for a $33$ monolayer thick film ($\sim 4.344$nm), $4$ $\Gamma$ subbands and bottom most
$X$ subband contribute, and this in turn affects the total carrier distribution, as shown in Fig. \ref{fig:conc_33AL}.
\subsection{Parabolic effective Mass and It's Validity}
A simple parabolic effective mass has been derived in this section at the minima of different
subbands to show some interesting transport properties of thin film. Parabolic effective mass
$m^*(i,j)$ for the $i^{th}$ valley and $j^{th}$ subband is defined as
\beq\label{eq:mstar}
m^*(i,j) = \frac{\hbar^2}{\partial^2E(i,j)/\partial {\bar{k}}^2(i,j)}
\eeq
\bfg[htbp!]
\vs{-1.5in}
\hs{-1.3in}
\includegraphics[scale=0.6]{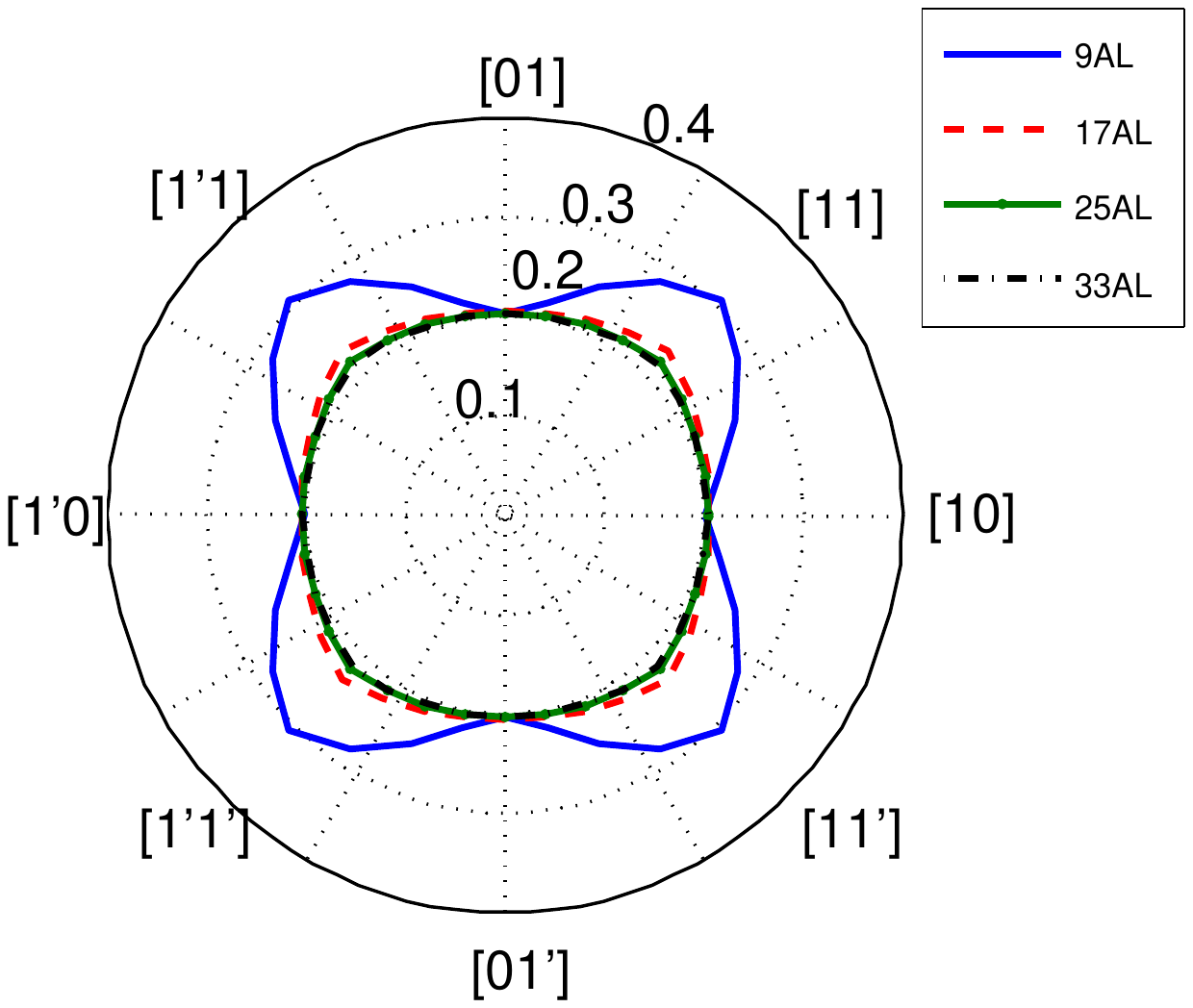}
\vs{-2.0in}
\caption{Variation of parabolic effective mass at the minimum of first
$\Gamma$ subband with crystal direction and Si film thickness. Anisotropy
is observed at small film thickness.}\label{fig:Si_mstar_Gamma1}
\efg
\bfg[htbp!]
\vs{-1.8in}
\hs{-1.5in}
\includegraphics[scale=0.6]{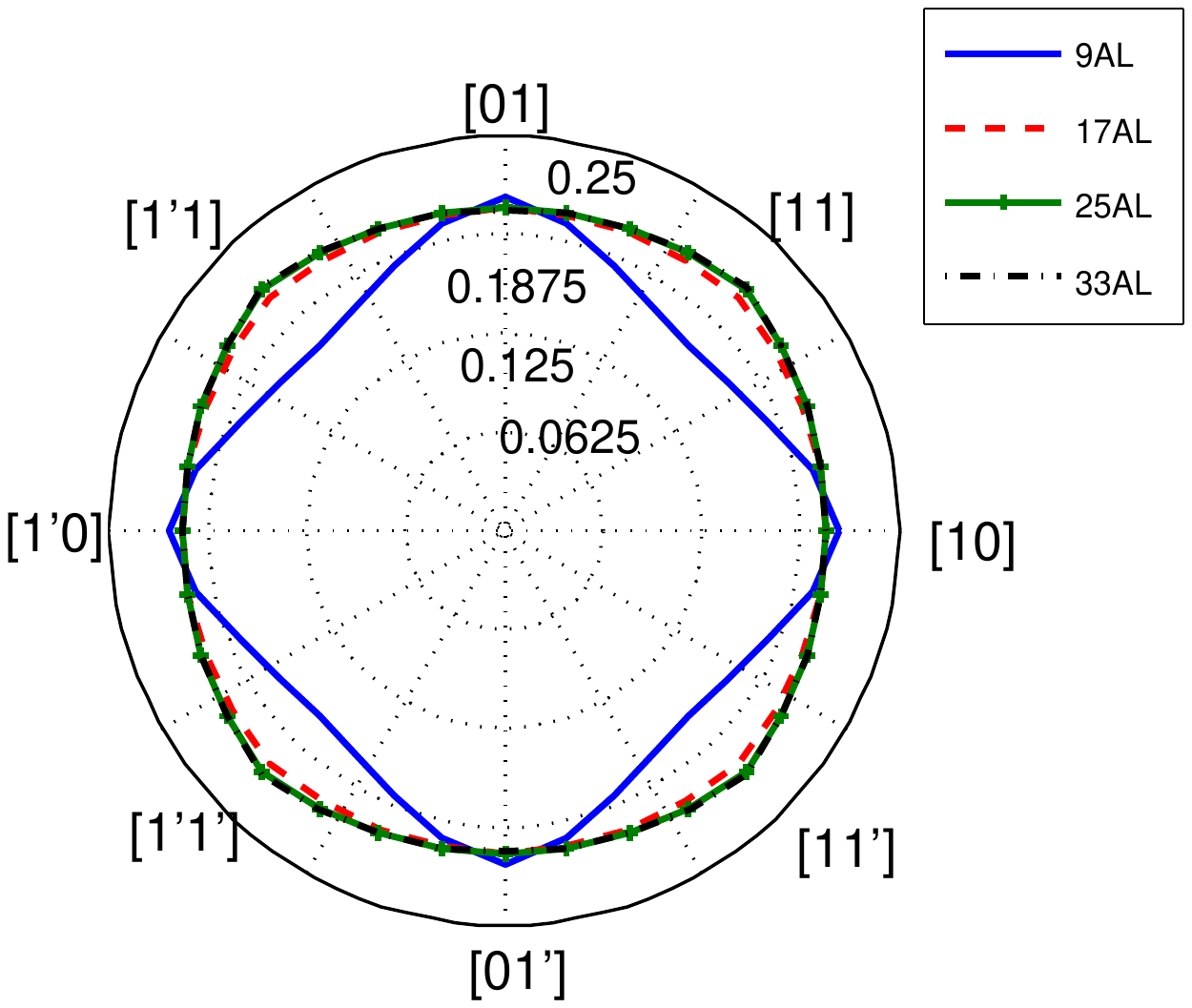}
\vs{-2.0in}
\caption{Variation of parabolic effective mass at the minimum of second
$\Gamma$ subband with crystal direction and Si film thickness. For very
small thickness ($\sim 1$nm), $<111>$ effective mass
is smaller than $<100>$ effective mass.}\label{fig:Si_mstar_Gamma2}
\efg
Fig. \ref{fig:Si_mstar_Gamma1} and \ref{fig:Si_mstar_Gamma2} show how parabolic
effective mass ($m^*$), normalized to electron rest mass ($m_0$), calculated at the minima of
two bottom most subbands $\Gamma_1$ and $\Gamma_2$ vary along different crystal direction,
for four different film thicknesses. One should note that, in both cases, for $9$ monolayer thick film,
effective mass is highly anisotropic. For $\Gamma_1$, effective mass increases as one moves
from $[10]$ direction to $[11]$ direction, whereas, it reduces for $\Gamma_2$ valley.
However, for larger thickness, effective mass in both the valleys, becomes fairly isotropic.
\bfg[htbp!]
\vs{-1.8in}
\hs{-0.7in}
\includegraphics[scale=0.55]{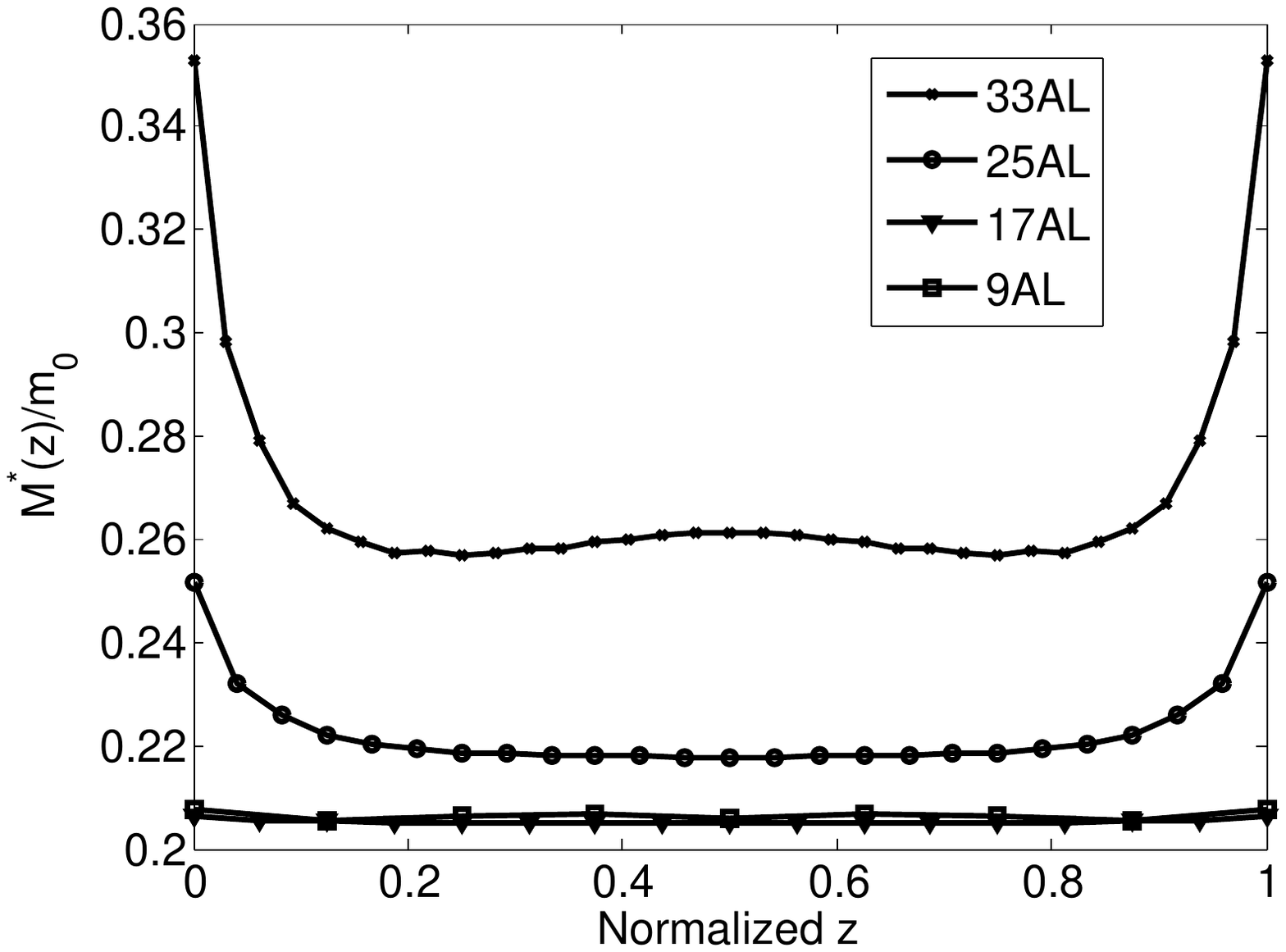}
\vs{-1.8in}
\caption{Distribution of $<100>$ $M^*(z)$ along film thickness
for various values of Si film thickness. Increasing thickness
increases $M^*(z)$ and reduces spatial uniformity.}\label{fig:effective_mass_distn}
\efg
Since the effective masses vary with subbands, and electron concentration in different
subbands has different spatial distributions, it is expected that effective mass should also
have a spatial distribution. A {\it`distributed effective mass'}, say $M^*(z)$, a function of
the depth $z$ along the thickness of the film, has been defined as:
\beq\label{eq:eff_mstar}
M^*(z) = \frac{1}{\sum_{i,j}\frac{W_{ij}(z)}{m^*(i,j)}}
\eeq
where $W_{ij}(z)$ represents the fractional contribution of electron concentration
at depth $z$ from $j^{th}$ subband of $i^{th}$ valley.
This way of defining $<100>$ $M^*(z)$ has the underlying assumption that
all the electrons (more generally, an equal fraction of electrons from each
subband of every valley) are moving along $<100>$ direction.
Fig. \ref{fig:effective_mass_distn} shows that for thinner
films, $<100>$ $M^*(z)$ is more or less uniform (which is because all the
electrons are in $\Gamma_1$ and $\Gamma_2$ subbands possessing almost same $<100>$
effective mass), but as thickness increases, the effective mass at position closer to surface becomes
larger than that of the central part of the film. Thus, for larger thickness,
carriers closer to center of the film are expected to be more mobile than those
which are closer to the surface. Note that, this effect is inherent to the intrinsic
film, coming from spatial distribution of wave functions associated with different subbands.
Another interesting observation is that, a $9$ monolayer thick film has (merginally)
larger $<100>$ effective mass than a $17$ monolayer thick one at all $z$. This is
because of the fact that the $\Gamma_2$ valley has larger effective mass for $9$ monolayer
thick film (Fig. \ref{fig:Si_mstar_Gamma2}).
\bfg[htbp!]
\vs{-1.5in}
\hs{-0.7in}
\includegraphics[scale=0.55]{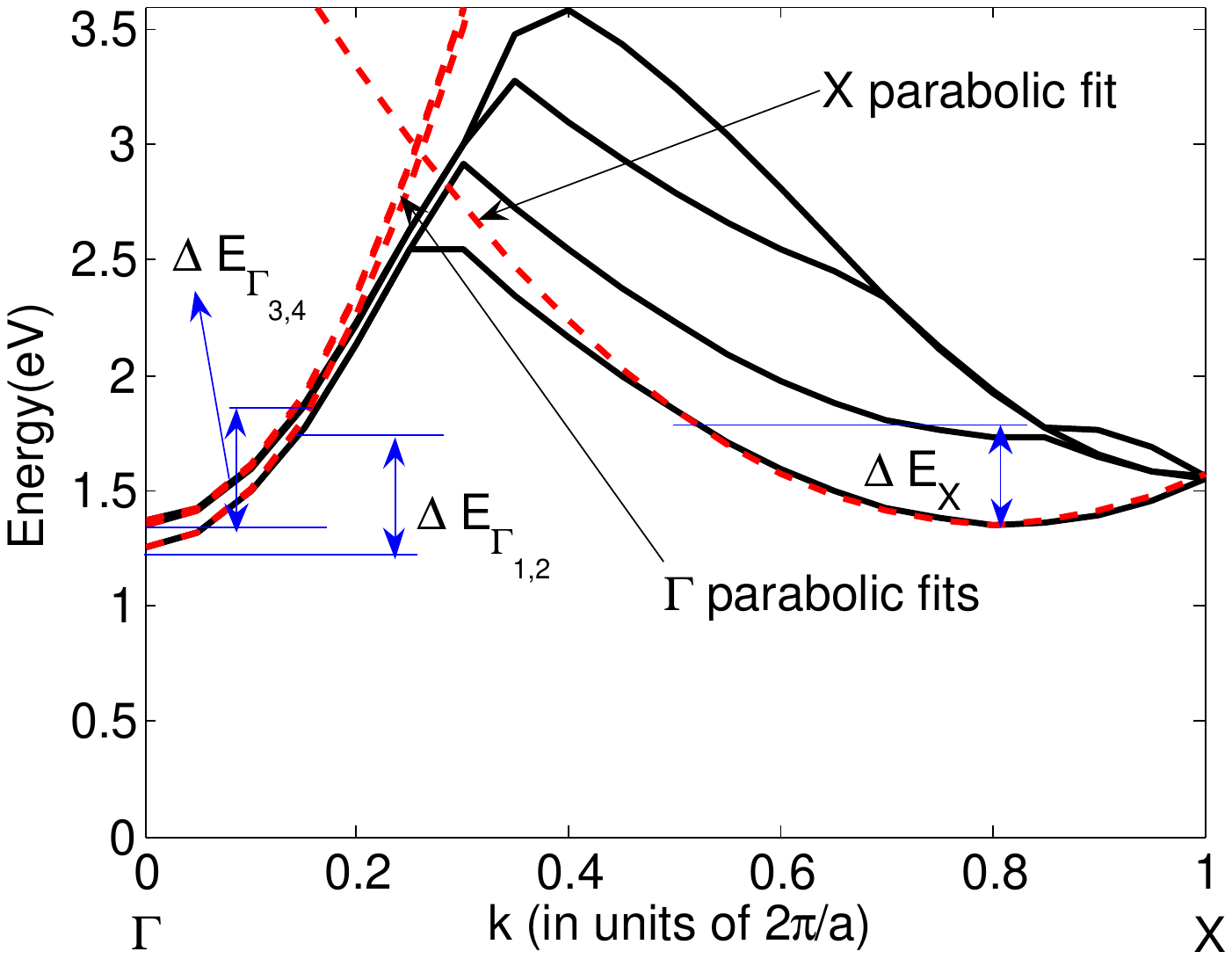}
\vs{-1.8in}
\caption{Original tight-binding bandstructure data and corresponding parabolic
fits at four bottom most $\Gamma$ subbands and bottom most $X$ subband for $<100>$ Si.
The fits are reasonable for energy values less than $\sim 0.5$eV referenced
from corresponding subband minima.}\label{fig:validity_mstar}
\efg

However, one should note that, parabolic effective mass approximation is
valid only for electrons with smaller energy in the conduction band.
The solid curves in Fig. \ref{fig:validity_mstar} show the $E-\bar{k}$
relationship along the $<100>$ direction for the four bottom most
conduction subbands, calculated from tight-binding method, as described in
sec. \ref{sec:bandcalc}. The dotted curves show the fitted parabolic bands with
same effective masses, as calculated from eqn. (\ref{eq:mstar}).
$\Delta E_{s}$ represents the electronic energy range in the
$s^{th}$ subband between
which the parabolic $E-\bar{k}$ tracks tight-binding $E-\bar{k}$ fairly well.
From Fig. \ref{fig:validity_mstar},
it is clearly
visible that parabolic bands fail to track the actual bands for electron
energies in excess of $\sim 0.5$eV, referenced from corresponding band
minimum, in all the cases. Nevertheless, the above simple analysis
gives good  qualitative insight about transport and mobility.
\section{Self-consistent Solution of Poisson-Schrodinger Equation}\label{sec:poiss_sch}
To study the effect of gate voltage on this structure, a model device has been assumed
consisting of a top gate and a bottom gate separated from the film by thin insulator
layers, as shown in Fig. \ref{fig:monolayers}. The Si film is assumed to be undoped.
Thus channel charge corresponds to only mobile charge. If $\phi(z)$ is the potential
at $z$, then one can write the 1-D Poisson equation as \cite{{yt00},{fs72}}
\beq\label{eq:poison2}
\frac{\partial^2\phi(z)}{\partial z^2} = \frac{qN}{\eps_0\eps_{rs}t}\left[ \sum_j\sum_{\bar{k}}2f(E_j^{\bar{k}})|\psi_j^{\bar{k}}(\phi,z)|^2 \right]e^{\frac{q\phi(z)}{k_BT}}
\eeq
where $q$ is electronic charge, $\eps_0$ is permittivity of vacuum and $\eps_{rs}$ is the relative
permittivity of the channel material.
The boundary conditions are derived from the fact that the normal component of the displacement vectors
inside Si and insulators will be the same at $z=0$ and $z=t$. Thus, at the boundaries, one can have
\beq\label{eq:bc1}
\eps_{rs} \frac{\partial \phi(z)}{\partial z}|_{z=0} = \eps_{r1} \frac{V_{g1} - V_{fb1} - \phi(0)}{t_{ox1}}
\eeq
and
\beq\label{eq:bc2}
\eps_{rs} \frac{\partial \phi(z)}{\partial z}|_{z=t} = \eps_{r2} \frac{V_{g2} - V_{fb2} - \phi(t)}{t_{ox2}}
\eeq
$\eps_{r1}$ and $\eps_{r2}$ are the relative permittivities of insulator1 and insulator2 respectively,
and, $t_{ox1}$ and $t_{ox2}$ are corresponding thickness of the insulators. $V_{fbi}$ is the flatband
voltage between the $i^{th}$ gate and channel.
Eqn. (\ref{eq:poison2}) can be solved iteratively to find $\phi(z)$.
To do this, first one assumes an initial potential profile $\phi(z)$, and then calculates
bandstructure, which in turn provides the correction to $\phi(z)$. This is iterated until it converges.
However, one should note that, in every iteration, one needs to calculate bandstructure. This is because the potential $\phi(z)$
adds a $z$ dependent perturbation to the crystal potential, and thus both $E_j^{\bar{k}}$ and $\psi_j^{\bar{k}}$
are function of $\phi(z)$. This makes the problem computationally intensive.

To reduce computation, the following approximation has been made. Note that, the potential $\phi(z)$
does not change drastically along $z$ (which is shown later), and thus, as far as change in bandstructure is concerned,
it's a fair assumption, that $\phi(z)$ is constant (=$\phi_c$) along $z$. Suppose, $H_0$ is the original
unperturbed $10N\times10N$ tight-binding Hamiltonian, $E_0$ and $\psi_0$ are the unperturbed eigen values and eigen
functions respectively. If one assumes that external potential only changes the
on-site energies, and not overlap integrals, then the perturbation $\Delta H$ can be written as
\beq
\Delta H = -q\phi_cI
\eeq
where $I$ is $10N\times10N$ diagonal unity matrix. Then,
\beq
H\psi_0 = (H_0 + \Delta H)\psi_0 = (E_0 - q\phi_c)\psi_0
\eeq
which means that all the energy eigenvalues will be shifted by same energy (in other words, no relative
change in energy eigenvalues), and the wave functions
remain in the unperturbed state. Thus, it is sufficient to calculate the bandstructure only once, and
the same $E_j^{\bar{k}}$ and $\psi_j^{\bar{k}}$ can be used through all the iterations. This reduces the
total runtime by nearly same number of times as the number of iterations it takes to solve the Poisson
equation (which varies roughly from $20$ to $200$ for different cases).

To validate the approximation, the amount of error being incurred
in the worst case (maximum gate voltage where band bending is maximum)
has been examined and the results are tabulated in Table \ref{tab:error},
for both $9$ and $33$ monolayer thick films. The error here has been
defined as $P_{error} = \frac{P_{approx} - P_{exact}}{P_{exact}}\times 100\%$ where
$P_{approx}$ is the approximate value of parameter $P$ and $P_{exact}$
is the exact value of the parameter.
\begin{table}
\bc
\begin{tabular}{|c|c|c|c|c|}
\hline
Thickness & \multicolumn{2}{|c|}{$\phi_{error}(\%)$} & \multicolumn{2}{|c|}{$n_{error}(\%)$}\\
\hline
(AL) & Mean & SD & Mean & SD\\
\hline
9 & -0.14 & 0.03 & 0.51 & 2.09\\
\hline
33 & 0.65 & 0.31 & 2.98 & 6.28\\
\hline
\end{tabular}
\caption{Mean and Standard deviation values of percentage error in
$\phi(z)$ and $n(z)$ for $9$ and $33$ monolayer thick films with
$V_g=1.5V$.}\label{tab:error}
\ec
\end{table}

To simplify the analysis, in the following, the metals
used as gate electrodes, have been assumed to have mid-gap work-function, and charge trapping inside insulators
is taken to be zero. This essentially means that the flatband voltages $V_{fb1}$
and $V_{fb2}$ are taken to be zero. For simulation, it has been assumed that
$t_{ox1}$ = $t_{ox2}$ = $1$nm, $V_{g1}$ = $V_{g2}$ and $\eps_{r1}$ = $\eps_{r2}$ = $3.9$.

\subsection{Volume Inversion}
Under the above mentioned assumption that bandstructure remains fairly the same under
application of gate voltage, one can write the carrier density distribution $n(z)$ as
\beq\label{eq:nz}
n(z) = n^0(z)e^{\frac{q\phi(z)}{k_BT}}
\eeq
where $n^0(z)$ is the intrinsic carrier density at $z$ and is given by
eqn. (\ref{eq:n0}) and $\phi(z)$ is the potential profile obtained by solving
the eqn. (\ref{eq:poison2}). Fig. \ref{fig:bias_conc_1V_9AL} and \ref{fig:bias_conc_1V_33AL} show the
distribution of carrier density over different subbands and the total density, for $9$
and $33$ atomic layer thick films respectively, when a $1$V supply has been applied to both the gates.
One should note the difference in shape of the carrier distribution as compared to
intrinsic case. Also, the $9$ atomic layer thick film shows a higher peak carrier
density compared to the $33$ atomic layer thick one, although the total integrated carrier
concentration is larger for thicker film. This can be explained
with {\it `potential pinning'} effect, as discussed later. However, at any $z$,
the fractional contribution from a subband to the total carrier density
remains the same at any applied gate voltage.
\bfg[htbp!]
\vs{-1.8in}
\hs{-0.7in}
\includegraphics[scale=0.55]{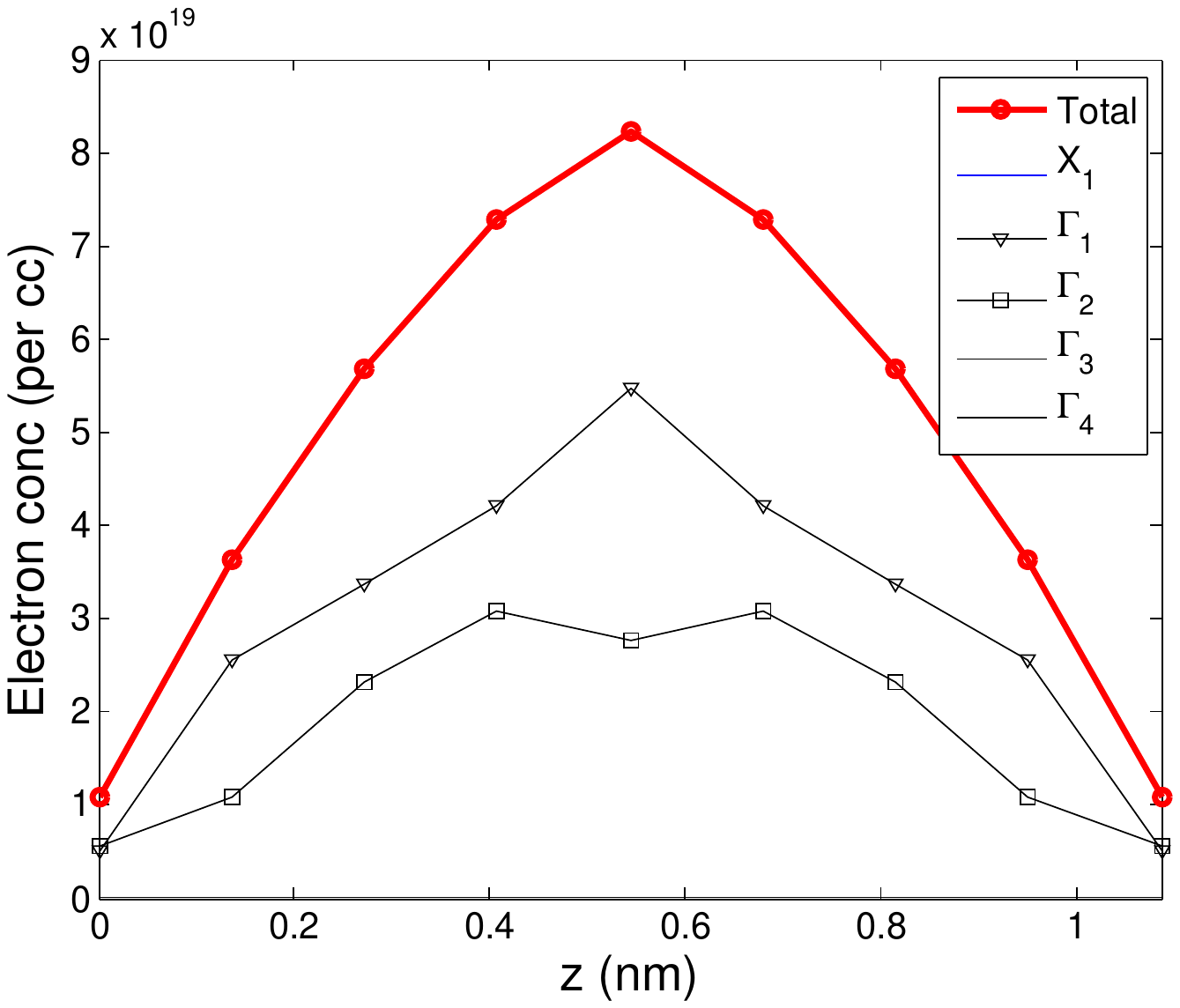}
\vs{-1.8in}
\caption{Total electron density distribution along the thickness and
contribution from different subbands for a $9$ atomic layer thick Si film
with $V_g$ = $1.0$V.}\label{fig:bias_conc_1V_9AL}
\efg
\bfg[htbp!]
\vs{-1.8in}
\hs{-0.7in}
\includegraphics[scale=0.55]{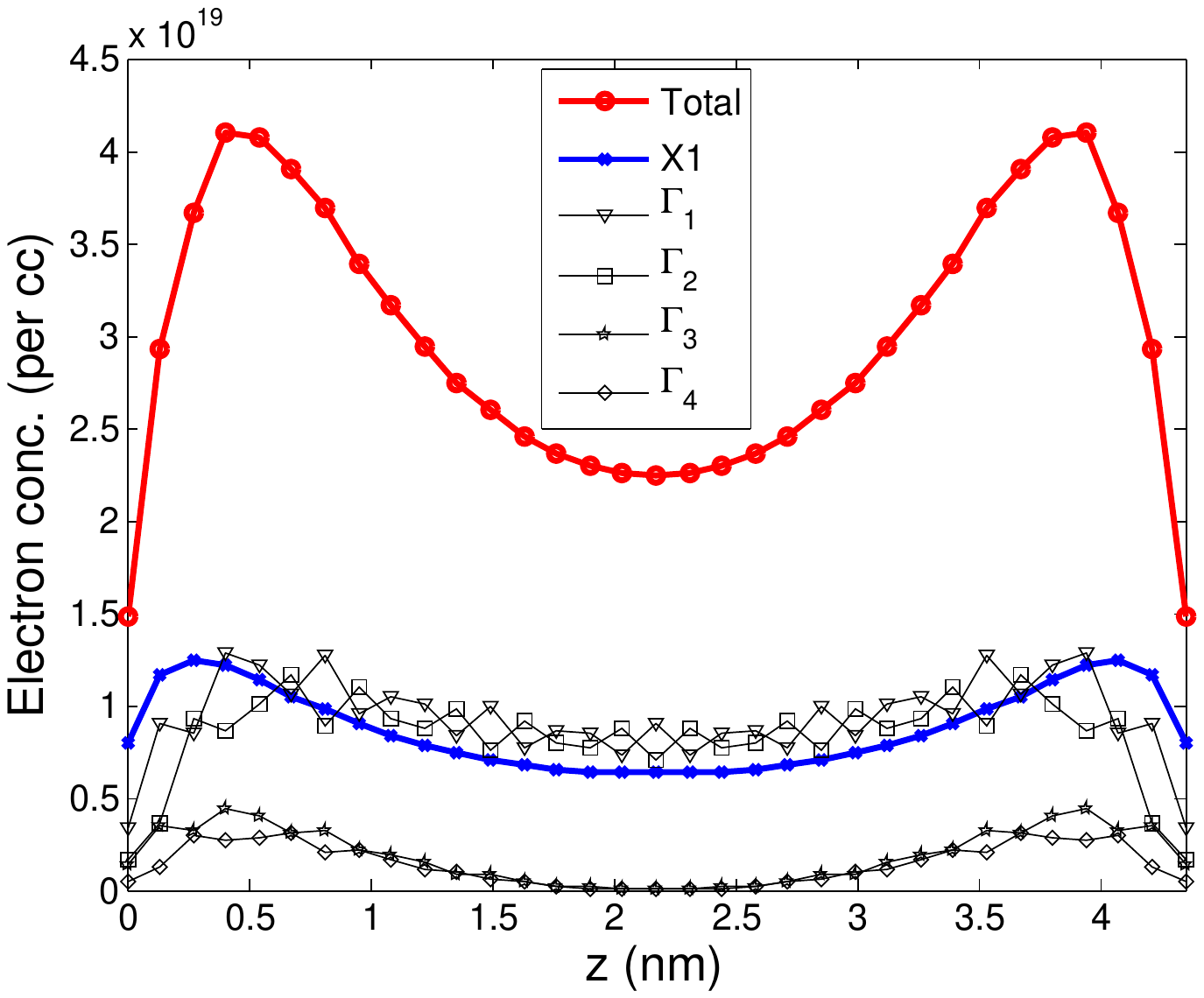}
\vs{-1.8in}
\caption{Total electron density distribution along the thickness and
contribution from different subbands for a $33$ atomic layer thick Si film
with $V_g$ = $1.0$V.}\label{fig:bias_conc_1V_33AL}
\efg
Fig. \ref{fig:normalized_carrier_9AL} and \ref{fig:normalized_carrier_33AL} show the normalized
carrier distribution along the depth $z$ at different gate voltages, for $9$ and $33$
atomic layer thick films. The corresponding potential distribution plot is shown in Fig. \ref{fig:potential_9_33}.
As gate voltage increases, the carriers of the {\it `central-peaked'} channel start spreading toward the
channel-insulator interface, and beyond
a particular threshold gate voltage $V_{vt}$, the carrier distribution will start showing two `humps' representing
separation of peaks of carrier concentration inside channel. $V_{vt}$ is defined by the condition
\beq
\frac{\partial^2n(z=\frac{t}{2})}{\partial z^2} \mid_{V_g = V_{vt}} = 0
\eeq
$V_{vt}$ for a thinner film is expected to be larger than a thicker one. In other words, carriers try to stay closer to
the center for thinner film. With increase in gate voltage,
the shifting of carrier density peaks toward the gates can be thought of as a {\it `carrier pulling effect'}
of the gate voltage. The effect can be explained from the potential profile $\phi(z)$ which
is smaller at points closer to the center of the film. From eqn. (\ref{eq:nz}), one notices
that the overall carrier density profile arises from
the individual contribution of the two terms $n^0(z)$ and $e^{\frac{q\phi(z)}{k_BT}}$.
$n^0(z)$ peaks at the center of the film, and reduces toward the surface whereas $e^{\frac{q\phi(z)}{k_BT}}$
follows the opposite trend.
Thus, at sufficiently large voltage, it is possible that the peak of carrier density occurs at the surface,
which is qualitatively same as the classical picture, where the exponential term dominates so much that
the effect of `quantum mechanical' distribution of intrinsic carrier density gets nullified.
\bfg[htbp!]
\vs{-1.8in}
\hs{-0.7in}
\includegraphics[scale=0.55]{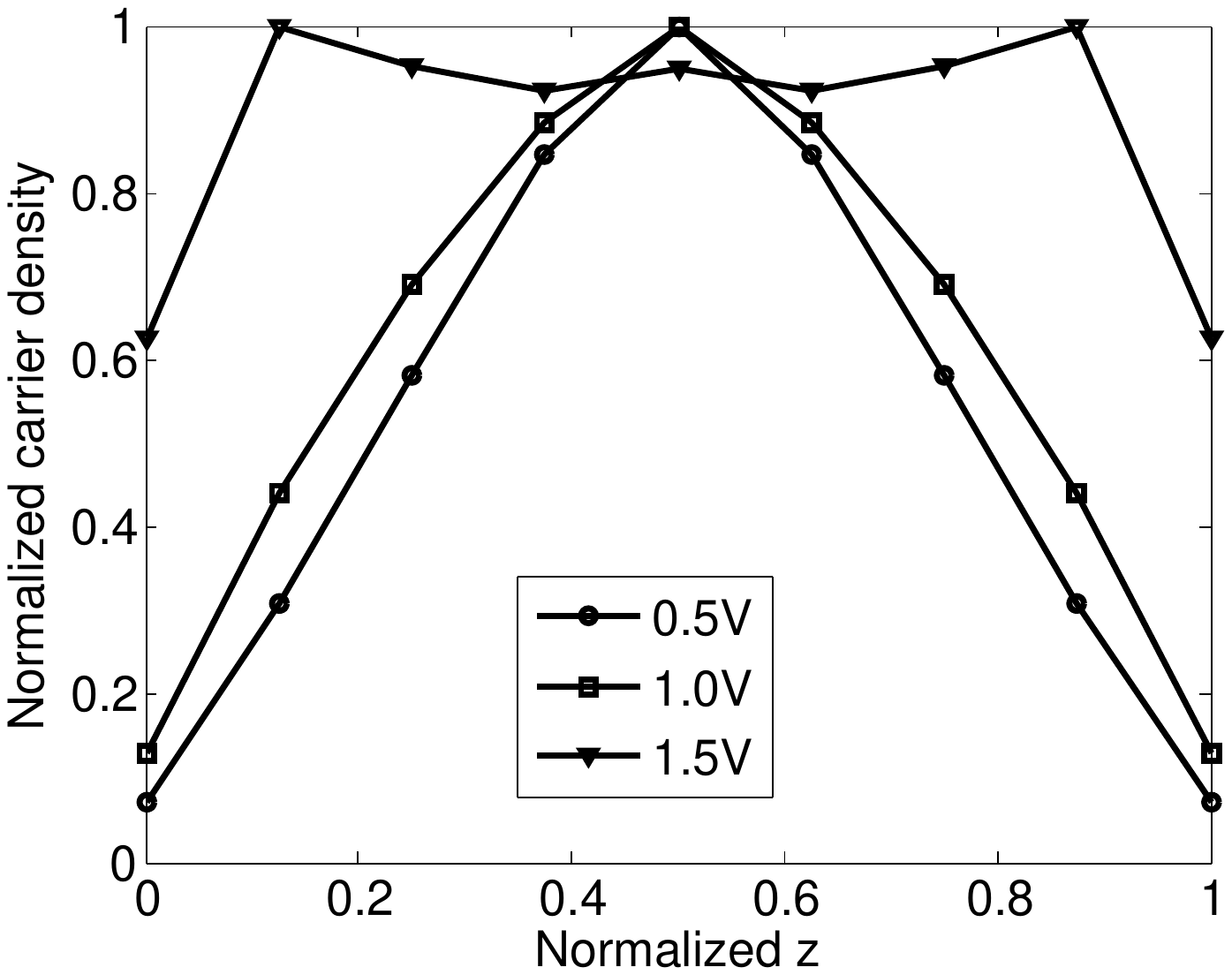}
\vs{-1.8in}
\caption{Normalized total electron density distribution
along the film thickness for a $9$ atomic layer thick Si film
with $V_g$ = $0.5$V, $1.0$V and $1.5$V. Single peak is
observed even at larger gate voltages.}\label{fig:normalized_carrier_9AL}
\efg
\bfg[htbp!]
\vs{-1.8in}
\hs{-0.7in}
\includegraphics[scale=0.55]{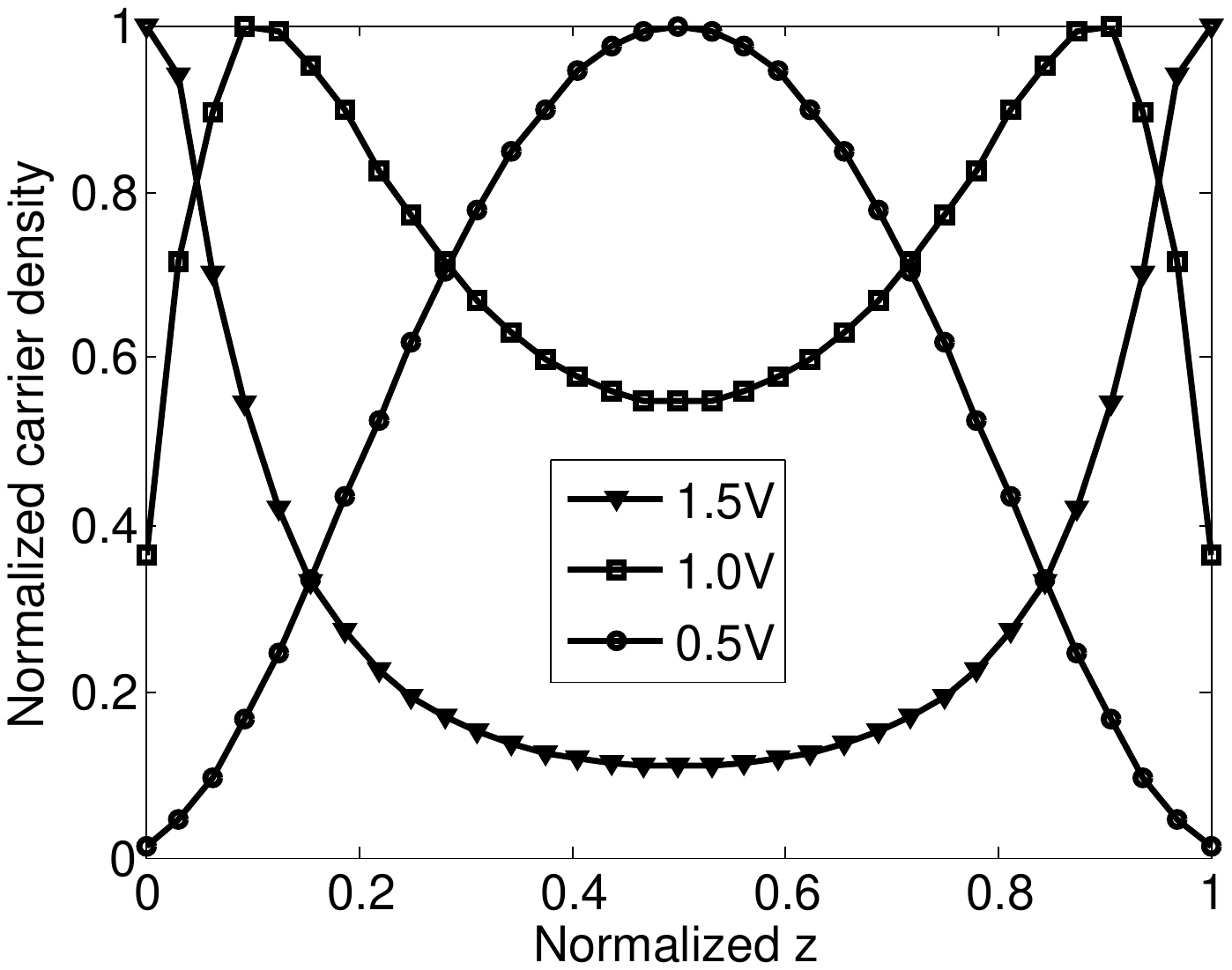}
\vs{-1.8in}
\caption{Normalized total electron density distribution
along the film thickness for a $33$ atomic layer thick Si film
with $V_g$ = $0.5$V, $1.0$V and $1.5$V. `Double hump'
characteristics is clear at higher gate voltages.}\label{fig:normalized_carrier_33AL}
\efg
As evident from Fig. \ref{fig:potential_9_33}, the potential profile for thinner films is more
uniform, and actually `pinned' at a higher value compared to the films of larger thickness leading
to higher peak carrier concentration in thinner films as shown in
Fig. \ref{fig:bias_conc_1V_9AL} and \ref{fig:bias_conc_1V_33AL}.
\bfg[htbp!]
\vs{-1.8in}
\hs{-0.7in}
\includegraphics[scale=0.55]{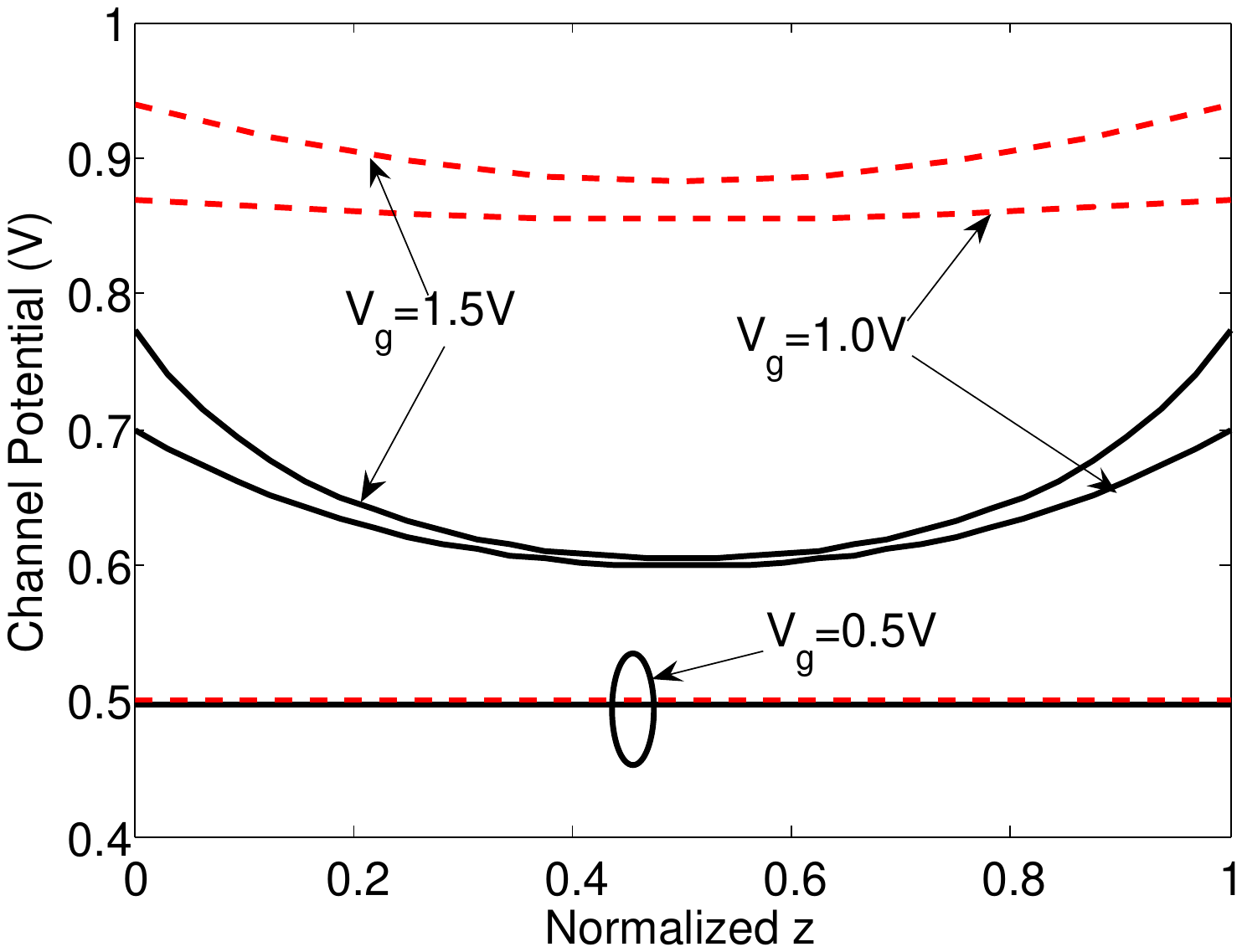}
\vs{-1.8in}
\caption{Potential distribution $\phi(z)$ along the thickness of thin Si film
with three different gate voltages: $0.5$V, $1.0$V and $1.5$V.
The solid and dotted curves represent $33$ and $9$ monolayer
thick films, respectively.}\label{fig:potential_9_33}
\efg

The above explanation becomes even more clarified from Fig. \ref{fig:phi_s_phi_0}.
Initially, for small gate voltage, both the surface potential $\phi_s$ and film center
potential $\phi_0$ increase simultaneously at the same rate, and thus
there will be only a single channel whose peak is at the center of the film.
Beyond a certain gate voltage, $\phi_s$ and $\phi_0$ bifurcate,
and $\phi_0$ saturates very quickly. However, $\phi_s$ keeps increasing, though at a
much slower rate than earlier, causing higher carrier concentration at points closer to the surface,
finally destroying the single peaked channel and creating two channels of `double hump'
shape. Note that, for $9$ atomic layer thick film,
the {\it `pinning'} voltages are higher than $33$ atomic layer film.
\bfg[htbp!]
\vs{-1.8in}
\hs{-0.7in}
\includegraphics[scale=0.55]{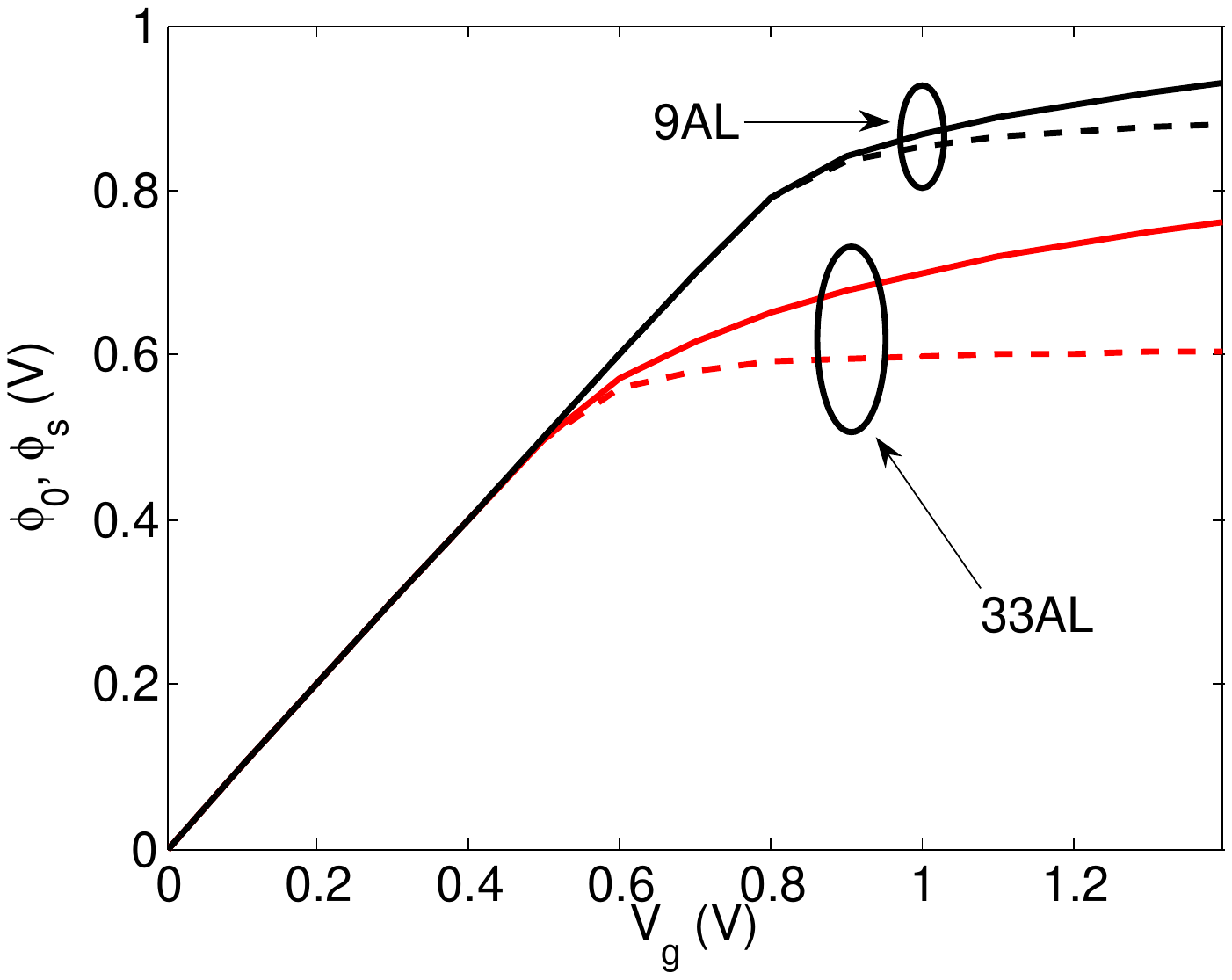}
\vs{-1.8in}
\caption{Variation of surface potential $\phi_s$ (solid lines) and film center
potential $\phi_0$ (dotted lines) with applied gate voltage $V_g$, for two
different film thicknesses.}\label{fig:phi_s_phi_0}
\efg
\subsection{Channel Charge Distribution}
The {\it `carrier pulling effect'} explained in the previous section can have immense
impact on the spatial distribution of total channel charge and hence device performance. Consider
a DGFET where the source end is grounded and the drain is connected to $V_{DD}$ which is same as
applied gate voltage $V_g$. At any point $x$ along the channel (with $x=0$ being taken as the source end),
suppose the quasi-Fermi level
is $V_{qf}(x)$. $V_{qf}(x)$ can be assumed as independent of $z$. Then, the Poisson equation in
eqn. (\ref{eq:poison2}) gets modified as \cite{yt04}
\begin{eqnarray}\label{eq:poison3}
\frac{\partial^2\phi(x,z)}{\partial z^2} & = & \frac{qN}{\eps_0\eps_{rs}t}\left[ \sum_{j}^{}\sum_{\bar{k}}^{}2f(E_j^{\bar{k}})|\psi_j^{\bar{k}}(\phi,z)|^2\right]
\nonumber \\
& & \times e^{\frac{q(\phi(x,z) - V_{qf}(x))}{k_BT}}
\end{eqnarray}
where all references have been made from grounded source chemical potential.
The carrier density $n(x,z)$ should now be a function of $V_{qf}(x)$ as well,
which in turn depends on the drain voltage. Deriving the exact drain current
for a nano-MOSFET needs proper attention on transport model and the details
will be communicated separately. However, to get a quantitative estimate,
a long channel device ($L=1\mu m$, $W=1\mu m$)
with constant mobility has been assumed. The drain current and $V_{qf}(x)$ have been
calculated using a similar procedure as in \cite{yt04}, by self-consistently
solving eqn. (\ref{eq:poison3}) with drain current continuity equation.
The extracted normalized carrier concentration $n'(x,z)$ is plotted over the whole
channel region in Fig. \ref{fig:channel_charge_distn} for different cases. At any $x=x_0$,
$n'(x_0,z)$ has been defined as
\beq
n'(x_0,z) = \frac{n(x_0,z)}{MAX_z\{n(x_0,z)\}}
\eeq
where $MAX_z\{.\}$ represents the maximum value of $\{.\}$ over $z$.
The normalization is done in such a way which clearly shows the spatial shape of carrier
distribution at different $x$.
It is observed that to the source end ($x=0$), due to larger potential difference between gate and channel, there clearly
exist two distinct carrier density peaks. However, as one moves toward the drain end, the potential difference between gate and channel reduces,
and the two distinct peaks merge together producing a single center-peaked channel.
Also, The magnitude of the total carrier concentration reduces toward the drain end.
Putting in another way, near the source, carriers
stay closer to the surface, and near the drain, carriers stay closer to the center of the channel. Thus, near the source,
one expects more surface scattering and away from it, surface scattering is expected to reduce. Similar effects are true
for gate leakage and gate capacitance, which can no longer be assumed to be uniform along the channel.
It is evident from  Fig. \ref{fig:channel_charge_distn} that this effect is more pronounced in thicker
channel devices, and at higher operating voltages. If the channel is thin enough, as is the case of (b2) in
Fig. \ref{fig:channel_charge_distn}, it is possible to have a single peaked channel all over the device.
\bfg[htbp!]
\vs{-1.8in}
\hs{-0.7in}
\includegraphics[scale=0.55]{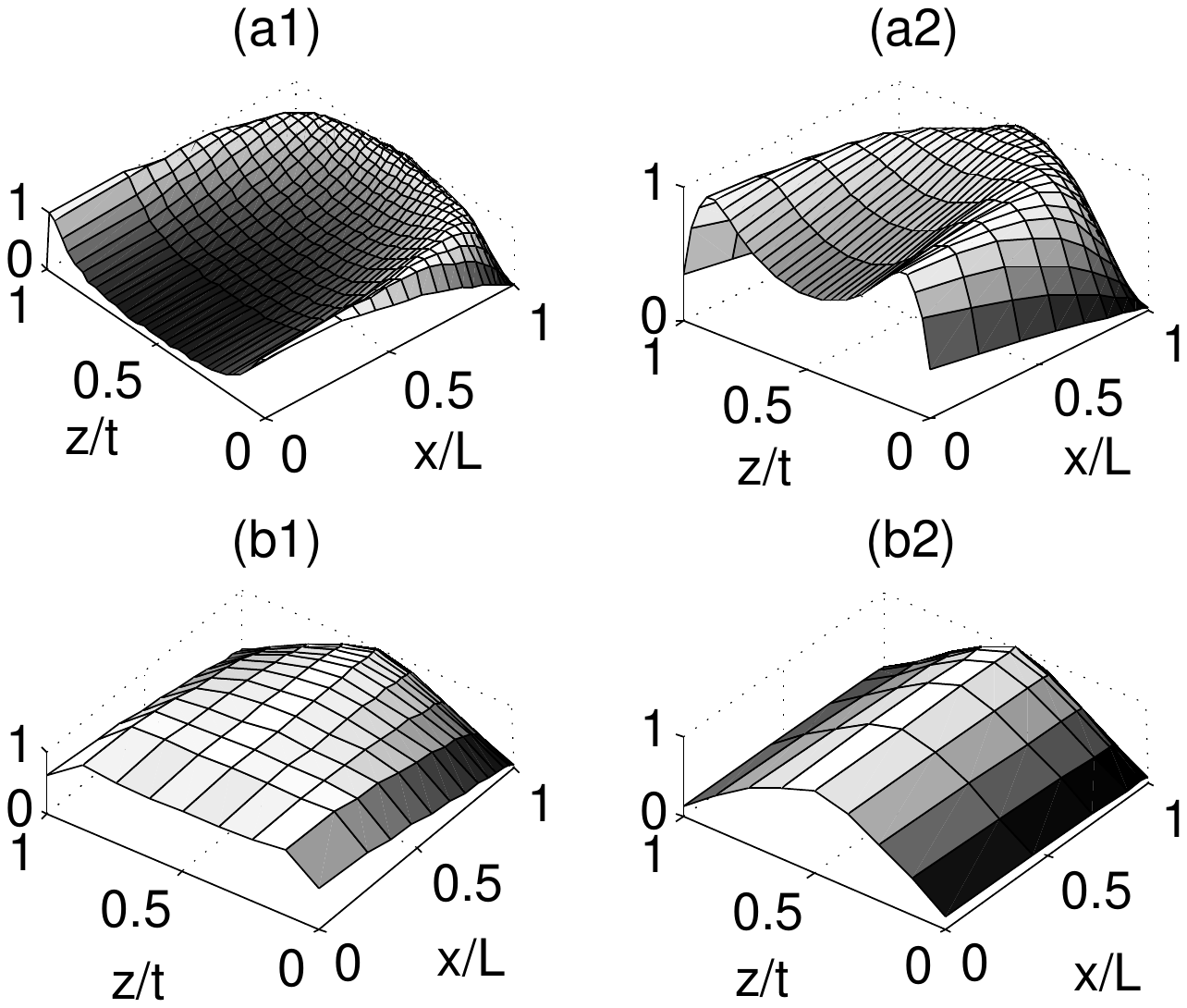}
\vs{-1.8in}
\caption{Normalized carrier distribution over the whole Si channel
for four different cases: (a1) $33$ monolayer thick channel, $V_g=1.5$V, $V_{DD}=1.5$V
(a2) $33$ monolayer thick channel, $V_g=1.0$V, $V_{DD}=1.0$V
(b1) $9$ monolayer thick channel, $V_g=1.5$V, $V_{DD}=1.5$V
(b2) $9$ monolayer thick channel, $V_g=1.0$V, $V_{DD}=1.0$V.}\label{fig:channel_charge_distn}
\efg

\subsection{Evolution of Effective Mass Along DGFET Channel}
In sec. \ref{sec:intrinsic}, it has been discussed in detail
how $<100>$ $M^*(z)$ varies along the thickness for a thin film Si.
Now, when one considers the channel of a DGFET, as has been
discussed in the previous section, the potential difference
between gate and channel changes from source end to drain end.
Thus total number of carriers at different subbands also changes along
the channel. Keeping this in mind, one can define an {\it `average
effective mass'}, $M_e^*(x)$
\beq\label{eq:Mex}
M_e^*(x) = \frac{\int_0^t n(x,z) dz}{\int_0^t \frac{n(x,z)}{M^*(z)} dz}
\eeq
which is basically harmonic average over the carriers along the thickness
at a particular position $x$ along channel length. Physically, this indicates
the `average' effective mass of an electron
located at distance $x$ from the source along the channel. Strictly speaking,
eqn. (\ref{eq:Mex}) is valid only under the assumption that
the vertical field is fairly constant and each electron suffers same
scattering rate. Although this is not a very good approximation,
but it gives an idea of how the channel charge distribution can affect
spatial distribution of carrier effective mass.
\bfg[htbp!]
\vs{-1.8in}
\hs{-0.7in}
\includegraphics[scale=0.55]{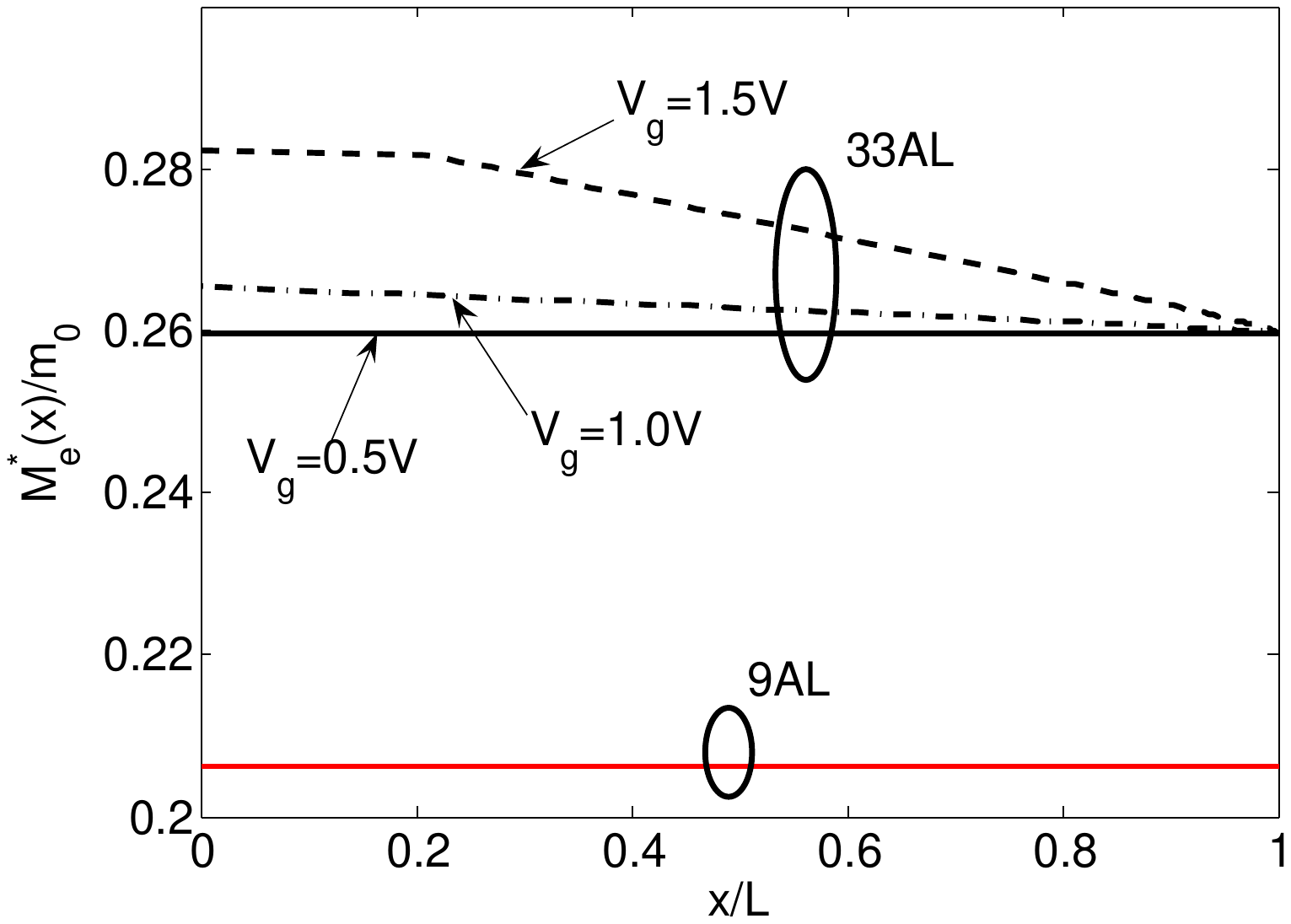}
\vs{-1.8in}
\caption{Variation of {\it `average effective mass'} $M_e^*(x)$
of carriers with channel along $<100>$ direction
for different channel thickness values and different gate voltages $V_g$
with $V_{DD}=V_g$.}\label{fig:mstar_evolution}
\efg
Fig. \ref{fig:mstar_evolution} shows for very small thickness channel (e.g. $9$ monolayer thick),
$M_e^*(x)$ hardly varies with $x$ as well as $V_g$. However, for higher thickness, a gradual
decrease in $M_e^*(x)$ is observed from source end to drain end, and the effect is more
prominent for higher gate voltages.
\subsection{Channel Capacitance}
Qualitatively, compared to classical analysis, the charge distribution in a
quantum analysis, has two major differences: 1) The total charge in the channel
reduces and 2) The charge distribution peak shifts from the surface toward the
center of the film. The extent of the shift depends on the applied gate voltage.
Both these effects cause a change in total gate capacitance \cite{yt01}.
The channel capacitance per unit volume $C_{si}(z)$ at a depth $z$ can be defined as the rate of
change of charge per unit volume with respect to the potential at that point. Mathematically,
\beq
C_{si}(z) = \frac{\partial Q_{si}(z)}{\partial \phi(z)}
\eeq
where $Q_{si}(z)$ is given by
\beq
Q_{si}(z) = \frac{qN}{t} \left[ \sum_j\sum_{\bar{k}}2f(E_j^{\bar{k}})|\psi_j^{\bar{k}}(\phi,z)|^2 \right]e^{\frac{q\phi(z)}{k_BT}}
\eeq
\bfg[htbp!]
\vs{-1.8in}
\hs{-0.7in}
\includegraphics[scale=0.55]{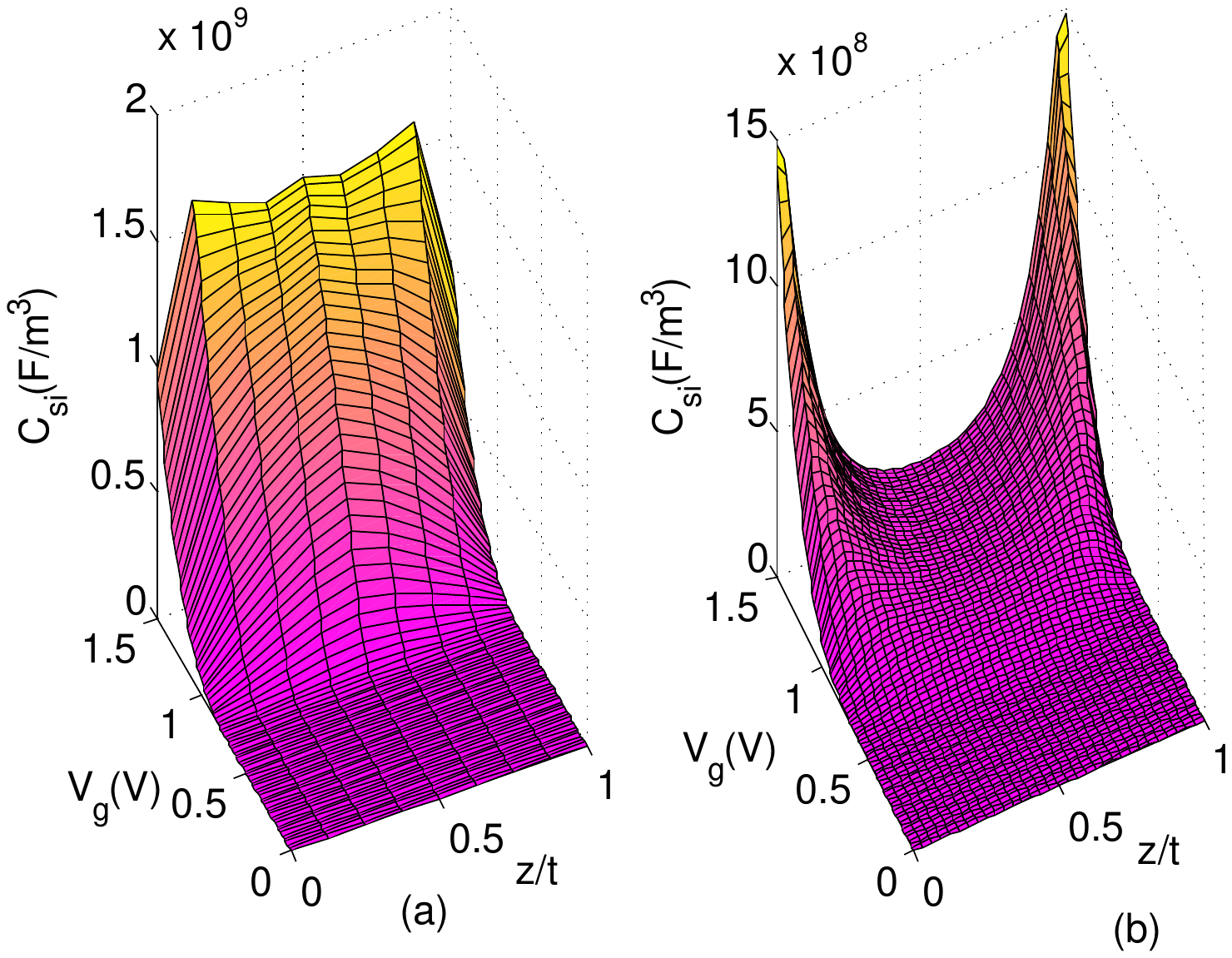}
\vs{-1.8in}
\caption{$C_{si}$ as a function of $V_g$ and depth $z$ along thickness of
channel for (a) $9$ monolayer thick channel and (b) $33$ monolayer
thick channel.}\label{fig:Csi}
\efg
Fig. \ref{fig:Csi} shows the variations of channel capacitance with $z$ and $V_g$. One should
note that the position $z=z_{max}$ inside the film, where the coupling with gate is maximum, varies with
gate voltage and as gate voltage increases, $z_{max}$ shifts toward the surface.
\section{Conclusion}\label{sec:conclude}
A detailed analysis of generic ultra-thin-body DGFET has been performed in this work.
The channel material has been chosen to be Si, but the analysis and
methodology can be readily extended to other promising channel materials as well.
Ultra thin film of Si has been shown to have larger and direct band gap,
as opposed to bulk. It has also been shown that the intrinsic carrier
concentration is not only less compared to bulk, but also has a distribution
over the channel thickness, peaking at the center. The contributions of different
subbands from different valleys to both intrinsic carrier concentration as well as
effective mass have been analyzed. The spatial distribution of distributed $<100>$
effective mass along thickness has been studied. It has also been shown that along $<100>$
direction, parabolic effective mass is fairly valid till an electronic
energy of $\sim 0.5$ eV from the corresponding subband minima in all
the relevant valleys. A detailed insightful analysis of volume inversion
has been performed and using {\it `carrier pulling effect'} of gate voltage,
channel charge distribution in a DGFET has been predicted.
The effects of channel charge distribution on effective mass and channel
capacitance have been analyzed critically.

\end{document}